\documentclass[10pt,journal,compsoc]{IEEEtran}
\usepackage{amsmath,amsfonts}
\usepackage{array}
\usepackage{textcomp}
\usepackage{stfloats}
\usepackage{braket}
\usepackage{url}
\usepackage{verbatim}
\usepackage{graphicx}
\usepackage{mathtools}
\usepackage{amsmath,amsfonts}
\usepackage{array}
\usepackage{multirow}
\usepackage{array, caption}
\usepackage{graphicx}
\usepackage{makecell}
\usepackage{amssymb, nccmath}
\usepackage{textcomp}
\usepackage{url}
\usepackage{subfigure}
\usepackage{tikz}
\usepackage{pgfplots}
\usepackage{tikz}
\usepackage[numbers,sort&compress,square]{natbib}
\usepackage{ragged2e}
\usepackage[utf8]{inputenc}
\usepackage{tikz}
\usetikzlibrary{trees}
\usetikzlibrary{shapes}
\usetikzlibrary{arrows,calc,positioning}
\usepackage{pgfplots}
\usepackage{graphicx}
\usepackage{textcomp}
\usepackage[absolute,overlay]{textpos}
\usepackage{comment}
\DeclareUnicodeCharacter{FB01}{fi}
\usepackage{graphicx}
\usepackage{caption}
\usepackage{csquotes}
\usepackage{colortbl}
\usetikzlibrary{arrows,shadows} 
\usepackage{tikz}
\usetikzlibrary {positioning}
\usetikzlibrary{trees}
\usetikzlibrary{arrows}

\usepackage{amsmath,amsfonts}
\usetikzlibrary{patterns}
\definecolor{myhund}{HTML}{BE0032}
\definecolor{myfifty}{HTML}{FF003F}
\definecolor{mycolor1}{HTML}{F5F5DC}
\definecolor{mytwenty}{HTML}{8B008B}
\definecolor{myzero}{HTML}{FF007F}
\definecolor{pssfhun}{HTML}{0000FF}
\definecolor{pssffif}{HTML}{333399}
\definecolor{pssftwen}{HTML}{8A2BE2}
\definecolor{pssfzero}{HTML}{6699CC}
\definecolor{ff}{HTML}{D2691E}
\definecolor{bf}{HTML}{FFBF00}
\definecolor{rf}{HTML}{967117}
\definecolor{s1}{HTML}{CE2029}
\definecolor{s2}{HTML}{006D5B}
\definecolor{s3}{HTML}{8DB600}
\definecolor{r4}{HTML}{CE2029}
\usepackage{graphicx}

\usepackage{amsthm}
\usepackage{amssymb}
\usepackage[linesnumbered, ruled]{algorithm2e}
\makeatletter
\newcommand{\removelatexerror}{\let\@latex@error\@gobble}
\makeatother
\usepackage{dsfont}

\ifCLASSINFOpdf
\else
\fi

\makeatletter
\def\ps@IEEEtitlepagestyle{%
	\def\@oddfoot{\mycopyrightnotice}%
	\def\@oddhead{\hbox{}\@IEEEheaderstyle\leftmark\hfil\thepage}\relax
	\def\@evenhead{\@IEEEheaderstyle\thepage\hfil\leftmark\hbox{}}\relax
	\def\@evenfoot{}%
}

\def\mycopyrightnotice{%
	\begin{minipage}{\textwidth}
		\centering \scriptsize
		This article has been accepted in IEEE Transactions on Parallel and Distributed Systems Journal © 2023 IEEE. Personal use of this material is permitted. Permission from
		IEEE must be obtained for all other uses, in any current or future media, including reprinting/republishing this material for advertising or promotional purposes, creating new collective works, for resale or redistribution to servers or lists, or reuse of any copyrighted component of this work in other works. This work is freely available for survey and citation.
		
	\end{minipage}
}
\makeatother

\begin{document}
%
\title{ Performance Analysis of Machine Learning Centered Workload Prediction Models for Cloud}
%
%
%
%
\author{Deepika~Saxena,~Jitendra~Kumar,~Ashutosh~Kumar~Singh,~\IEEEmembership{Senior member IEEE},~and~Stefan~Schmid
	\IEEEcompsocitemizethanks{\IEEEcompsocthanksitem  D. Saxena is with Department of Computer Science, Goethe University Frankfurt, Germany. E-mail: 13deepikasaxena@gmail.com, 	d.saxena@em.uni-frankfurt.de\\
 J.Kumar is with the Department of Computer Applications, NIT Tiruchirappalli, Tamilnadu, India. E-mail: jitendra@nitt.edu\\
  A. K. Singh are with the Department of Computer Applications, NIT Kurukshetra. E-mail: ashutosh@nitkkr.ac.in\\
  S. Schmid is with TU Berlin, Germany, University of Vienna, Austria, and Fraunhofer SIT, Germany. E-mail: stefan.schmid@tu-berlin.de.  
}}

%
%

%

\markboth{IEEE TRANSACTIONS ON PARALLEL AND DISTRIBUTED SYSTEMS}{Shell \MakeLowercase{\textit{et al.}}: Bare Demo of IEEEtran.cls for Computer Society Journals}



\IEEEtitleabstractindextext{%
\begin{abstract}
The precise estimation of resource usage is a complex and challenging issue due to the high variability and dimensionality
of heterogeneous service types and dynamic workloads. Over the last few years, the prediction of resource usage and traffic has
received ample attention from the research community. Many machine learning-based workload forecasting models have been
developed by exploiting their computational power and learning capabilities. This paper presents the first systematic survey cum
performance analysis-based comparative study of diversified machine learning-driven cloud workload prediction models. The discussion
initiates with the significance of predictive resource management followed by a schematic description, operational design, motivation, and
challenges concerning these workload prediction models. Classification and taxonomy of different prediction approaches into five distinct
categories are presented focusing on the theoretical concepts and mathematical functioning of the existing state-of-the-art workload
prediction methods. The most prominent prediction approaches belonging to a distinct class of machine learning models are thoroughly
surveyed and compared. All five classified machine learning-based workload prediction models are implemented on a common
platform for systematic investigation and comparison using three distinct benchmark cloud workload traces via experimental analysis.
The essential key performance indicators of state-of-the-art approaches are evaluated for comparison and the paper is concluded by discussing the trade-offs
and notable remarks.


\end{abstract}

}

\maketitle

\IEEEdisplaynontitleabstractindextext

%
\IEEEpeerreviewmaketitle

\IEEEraisesectionheading{\section{Introduction}\label{sec:introduction}}

%
%
%
%
 \IEEEPARstart{T}{he} Cloud Computing (CC) paradigm empowered  with rapid elasticity, resource pooling, outsourced service management, broad network access, and pay-as-per-use model, facilitates scalable computing avenues with minimum upfront capital investment to enterprises, academia, research and all the stakeholders \cite{saxena2021op, saxena2021secure}.  CC is acting as a catalyst in driving business progress amidst growing uncertainty across the geographical boundaries by sustaining momentum and addressing the inconsistencies in global IT infrastructures \cite{mishra2022linking}. According to a recent survey report \cite{trend2022cloud}, it is anticipated that the global cloud computing market will reach USD 1,554.94 billion by 2030, registering a Compound Annual Growth Rate (CAGR) of 15.7\%. Moreover, all the emerging technologies including Internet of Things (IoT),  fog and edge computing, cyber-physical systems etc. emphatically depend on CC, because of their insufficiency of  storage and computing capabilities \cite{ren2022machine}.  
  
  \subsection{{Motivation}}
     Cloud Service Providers (CSP) employ virtualization \cite{saxena2023sustainable, song2013adaptive, saxena2020security, saxena2021osc, saxena2022ofp} of physical resources at datacentres to  maximize their revenue while serving the demand of computing instances with privilege of rapid scalability \cite{saxena2022high, saxena2021workload,  gupta2022quantum}. Therefore, the comphrehensive management of CC infrastructure entirely depends on the fine-grained provisioning of resources including storage, processing and networking etc. \cite{wang2022truthful, saxena2020communication, singh2021cryptography, saxena2022intelligent, xie2022random}.  The resource demands exhibit high variation over the time expediting over/under-utilization of physical machines, and Service Level Agreement (SLA) violation issues \cite{bi2019temporal}. {During peak load arrival, the aggregate demand of VM resources  exceeds the available resource capacity of the servers leading to overloaded servers and  performance degradation, for example, some VMs may crash, longer unavailability of resources and increased response time, etc. Whereas inadequate resource demands lead to the wastage of computational resources.  
     	 In order to manage the dynamic and random  requirement of resource capacities or handle over-/under-load, the migration of VMs in real time from an over-/under-loaded server to another server having sufficient resource capacity, leads to delayed execution. In this context, effective handling of incoming workloads via prior estimation is a prime requirement. An accurate prediction of load triggers reduction of resource wastage, minimum power consumption, and number of active servers by allowing only the required number of physical machines in active state. The precise information of workload imparts prior reservation of resources to execute and manage the forthcoming workload effectively, reduce response time, SLA violations, over-provisioning, and under-provisioning problems, and improve resource utilization, reliability, service availability. \cite{kabir2021uncertainty, saxena2020proactive, saxena2022fault, saxena2022intelligent1}}. 
    

\subsection{Workload Prediction Perspective}

\begin{figure}[!htbp]
	\centering
	\includegraphics[width=0.999\linewidth]{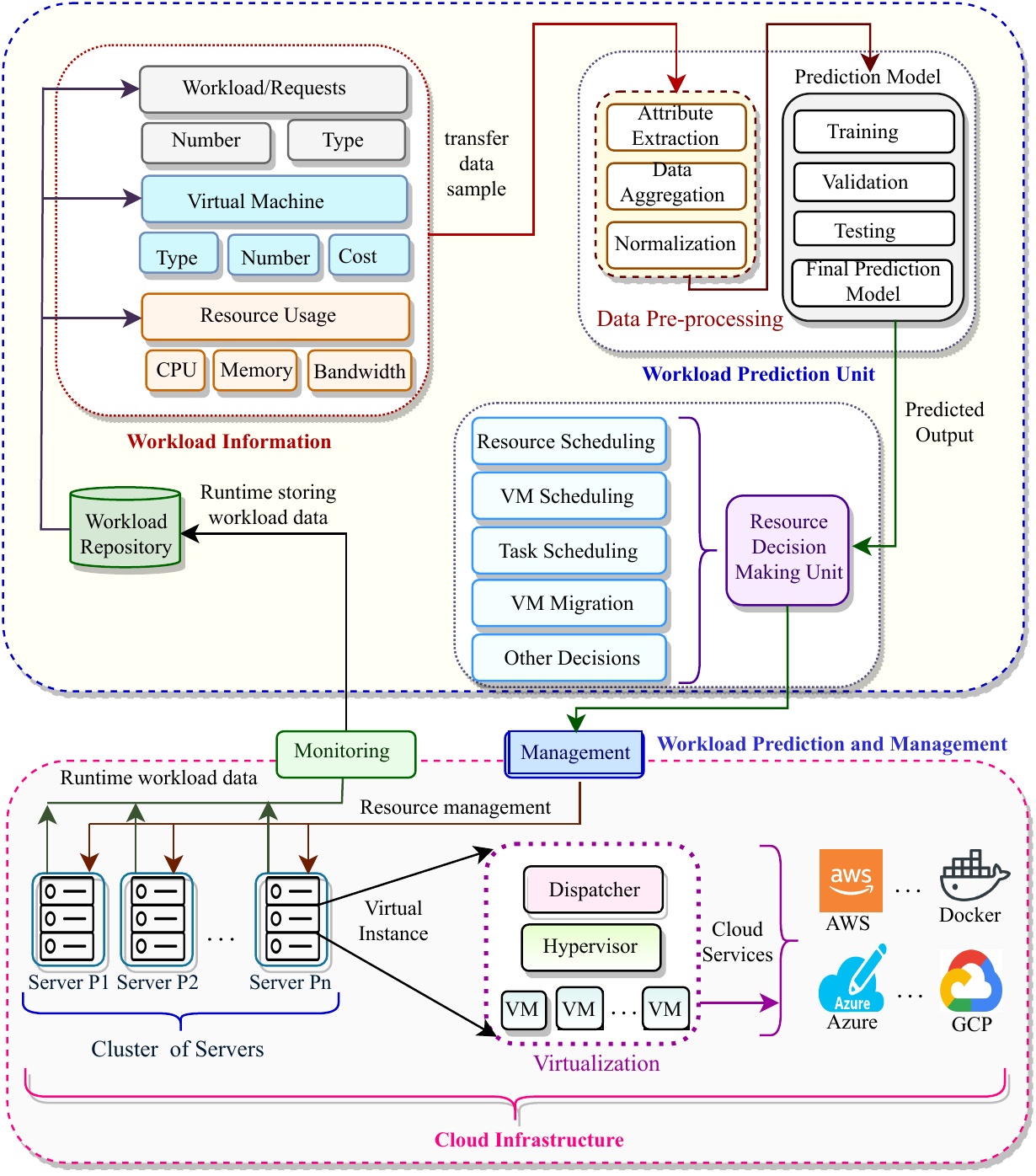}
	\caption{Schematic representation and application of Workload Prediction}
	\label{fig:workloadprediction}
\end{figure}

The perspective and utility of workload prediction for physical resource  management is illustrated in 
 Fig. \ref{fig:workloadprediction} via interactive information flow between the  `cloud infrastructure' and `workload prediction' blocks. The cluster of servers \{$ServerP_1$, $ServerP_2$, ..., $ServerP_n$\} ingrained with virtualization technology enable numerous virtual instances in the form of VMs, to cater services demands of cloud users. The virtualization allows sharing of physical machines among various applications (such as AWS, Docker, MS Azure, and GCP etc.) with the help of dispatcher and hypervisor. 
The resource usage information is monitored and recorded in a workload repository within workload prediction and management unit. The raw information including number and type of requests; number, type, and cost of VMs; resource (viz., CPU, memory, bandwidth) usage, is retrieved from the repository and transferred to workload prediction unit. The significant attributes from raw data samples are extracted, aggregated, and normalized during data pre-processing. A workload prediction model is employed which generates and evolves over series of stages such as training, validation, and testing for the real-time workload prediction. The final prediction model analyzes and estimates information regarding  resource usage, number and type of requests, etc. for rendering effective resource management decisions. The predicted resource information assists in assorting the needed physical resources proactively avoiding the run-time resource provisioning delay while satisfying Quality of Services (QoS) constraints. 

\subsection{{Research Challenges}}  
 {Assuredly, the cloud workload prediction plays an essential role in proactive auto-scaling and dynamic management of resources resulting into increased scalability and throughput of the systems, sustainability, fault-tolerance via proactive prediction of system failures.} However, there exists some major challenges addressing  the cloud workload prediction  which  are discussed as follows:
\begin{itemize}
	\item \textit{Heterogeneous Workload}: Cloud users submit different type of application requests, requiring heterogeneous resource capacities with varying priorities and pricing policies associated with their respective SLAs.	
	\item \textit{Uncertain Resource Demands}:
	The resource demand changes over time in an hour, day, week, month and years with respect to the type of workload and deadline of execution submitted by the user. Sometimes the traffic becomes bursty \cite{griner2021cerberus} which makes it difficult to estimate the upcoming resource demands and decide resource distribution.
	
	\item \textit{Dynamic Adaptation}: Since the cloud environment is highly variable and dynamic, it suffers from unexpected fluctuations, which put forth a crucial challenge of adaptability for workload prediction i.e., to adapt or re-generate in order to sustain and perform efficiently with the changing workloads. 
	
	\item \textit{Data granularity and Prediction window-size}: To decide the appropriate size of data sample or granule and length of prediction window i.e., for shorter or longer interval, is another critical challenge which directly effects learning of relevant patterns and developing correlations among extracted patterns. 		
\end{itemize}

\subsection{Paper Outline and Contributions}
This paper presents a comprehensive study of machine learning based cloud workload prediction models. The study begins in Section 1 with a discussion of the CC and the vital role and research motivation for the workload prediction  within CC environment. It is followed by a schematic representation with  an illustrative description of the  application of the load prediction and management in CDCs. Thenafter, research challenges depicting a commendable points of the  need and efficacious impact  of an accurate workload prediction for resource management and intervening critical issues are discussed. The  operational flow outlining the essential steps of workload prediction are rendered in Section 2. The intended research methodology is discussed in Section 3. This study aims to provide an extensive review of the most prominent and seminal machine learning based models proposed for the prediction of extensive range of cloud workloads. Accordingly, Section 4 discusses Evolutionary Neural Network  based prediction models,  Section 5 and Section 6 entail review of Deep Learning  and Hybrid learning based prediction models, respectively while Section 7 and Section 8 pertain to discuss Ensemble learning  and Quantum learning based prediction models, respectively.  Furthermore, the prediction models underlining the considered five categories are implemented on the common platform for evaluation and comparison of their performance in terms of various key performance indicators (KPI)s in Section 9. Finally, Section 10 concludes the study with a discussion of  trade-offs among the prediction models of different classes are remarked  with emerging research challenges addressing cloud workload forecasting along with their probable solution avenues are  discussed. To the best of the authors' knowledge, this is the first paper which aims to carry out a comprehensive experimental study on the machine learning based workload prediction models in the  context of resource management in CC. The key contributions of this paper are:
\begin{itemize}
	\item The commendable and recent cloud workload prediction models based on the machine-learning algorithms are designated with respect to their conceptual and operational characteristics into a  classification and taxonomical organization (Fig. \ref{fig:CloudComputing}).
	\item {To illustrate the generalized conceptual and operational design corresponding to the workload prediction approach belonging to each category, this paper figures out five specific machine learning model architectures and their working strategies.}
	   
	\item A critical discussion and comparison cosidering all the essential detail of state-of-the-art works   are provided and their
	features are analyzed to determine the future research scope addressing the limitation of the respective class based prediction model.

	\item {An implementation of the approaches associated to each of the five classes based prediction models on the same platorm is conducted for in-depth experimental analysis and comparison in terms of essential KPIs to  measure their performance followed by discussion of  trade-offs and notable remarks.}
\end{itemize}

 Table \ref{TableExp} gives the explanatory terms for the symbols, notations, abbreviations used throughout the manuscript.
\begin{table}[!htbp]
	\caption{Notations with their Explanatory Terms}
	\label{TableExp}
	\begin{center}
		\resizebox{0.49\textwidth}{!}{
			\begin{tabular}{|ll|ll|}\hline 
				\textbf{\textit{Notation}} & \textbf{\textit{Definition}} & \textbf{\textit{Notation}} & \textbf{\textit{Definition}}\\  \hline 
				$\mathcal{W}^{Ac}$ & actual workload & $\mathcal{X}$ & cell information \\ 
				$\mathcal{W}^{Pr}$ & predicted workload & $\mathcal{D}$ & Input data  \\ 
				$\mathcal{G}_{1}$ & first layer of LSTM & $\mathcal{B}$ & bias \\ 
				$\mathcal{CF}_{RU}$ & previous resource usage information & $\mathcal{WT}$ & weight matrix \\  
				$\mathcal{G}_{2}$ & sigmoid layer of LSTM & ${w}$ & neural weight\\
				$\textit{n}$ & number of input layer nodes & $\Theta$ & qubit\\
				$\textit{p}$ & number of nodes in hidden layer & $\mathcal{\uplus}$ & activation function\\
				$\textit{z}$ & number of base predictor (BP) & $\textit{y}^{In}$ & qubit input vector\\
				$\mathcal{MSE}$ & mean squared error & $\mathcal{MAE}$ & mean absolute error\\
				$\textit{m}$ & number of data samples & \textit{ENN} &evolutionary neural network \\
				\textit{CSP} & cloud service provider & \textit{CC} & cloud computing\\
				\textit{LSTM} & long short term memory & \textit{CDC} & cloud data centre\\
				\textit{RNN} & recursive neural network & \textit{DBN} & deep belief network\\
				\textit{Bi-LSTM}& bi-directional LSTM & \textit{DNN} & deep neural network\\

				\textit{SGD} & stochastic gradient descent & \textit{GRU} & gated recurrent unit\\
				\textit{PLR} & piecewise linear representation & \textit{LR} & logistic regression\\
				\textit{TSA} & top-sparse autoencoder & \textit{CP} & canonical polyadic\\
				\textit{OED} & orthogonal experimental design & \textit{S-G Filter} & savitzky-golay\\				
				\textit{PSO} & particle swarm optimization & \textit{BP} & back propagation\\
				\textit{ENN} & evolutionary neural network & \textit{QoS} & quality of service\\
				\textit{RMSE} & root mean squared error & \textit{MAE} & mean absolute error\\
				\textit{MAPE} & mean absolute percentage error & \textit{MSE} & mean squared error\\
				\textit{RMSSE} & root mean segment squared error & \textit{VM} & virtual machine\\
				\textit{MRPE}& mean relative predict error & \textit{DL} & deep learning\\
				\textit{CDF} & cumulative distribution frequency & \textit{RMSLE} & logarithmic RMSE\\		
				\textit{SaDE} & Self-adaptive differential evolution & \textit{AEF} & absolute error frequency\\
				\textit{BaDE} & bi-phase adaptive differential evolution & \textit{EQNN} & evolutionary QNN\\
				\textit{TaDE} & Tri-adaptive differential evolution & \textit{C-NOT} & controlled NOT gate\\
				\textit{ARIMA} & auto regressive integrated moving average & \textit{HL} & hybrid learning\\
				\textit{QNN} & quantum neural network & \textit{EL} & ensemble learning\\
				\hline 
		\end{tabular}}
	\end{center}
\end{table}

 \section{ Workload Prediction Operational Flow}
\begin{figure}[!htbp]
 	\centering
 	\includegraphics[width=1.0\linewidth]{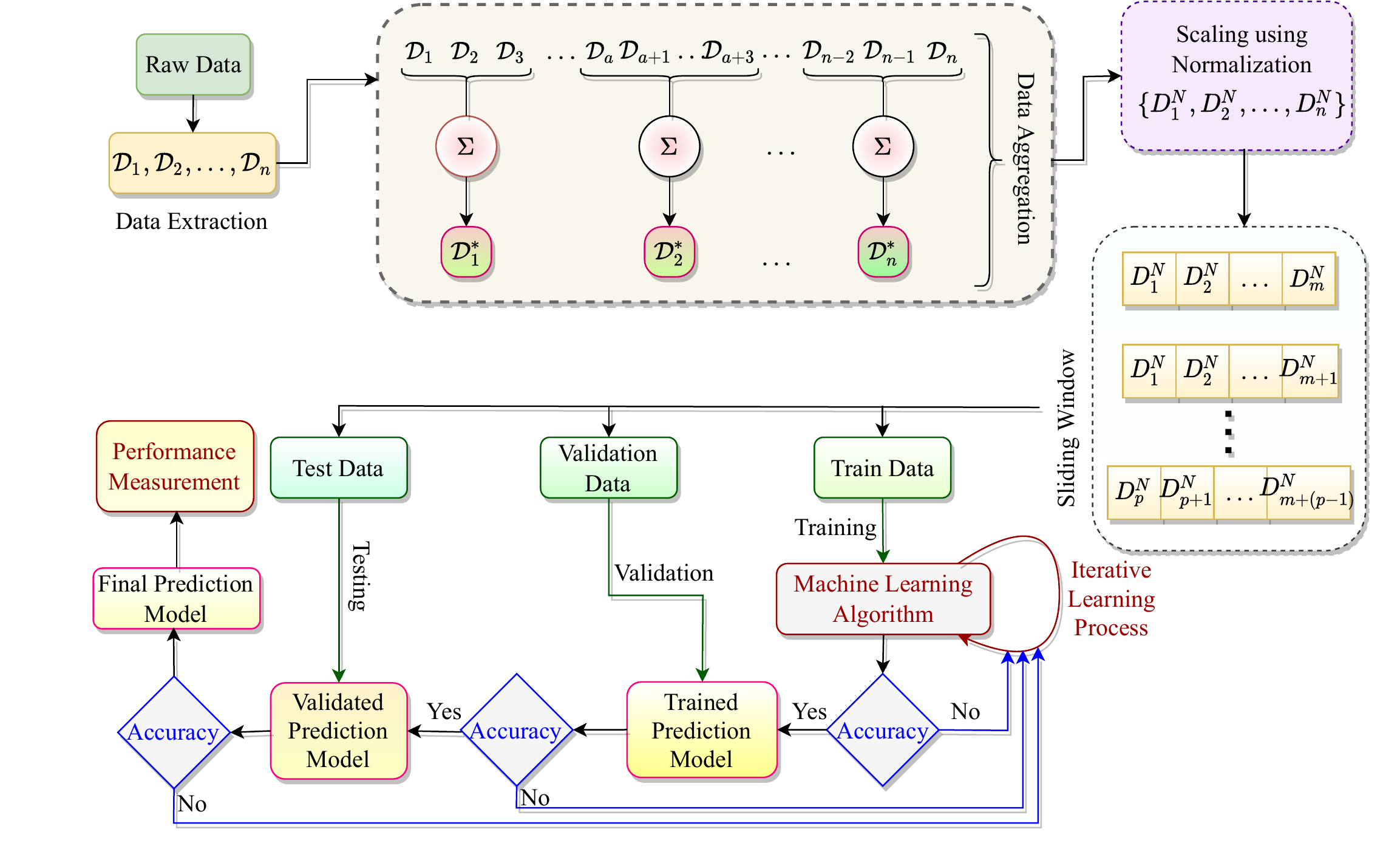}
 	\caption{Load prediction operational flow}
 	\label{fig:preprocessing}
 \end{figure}

 The essential steps intended for the workload prediction are outlined via illustration of an operational design in Fig. \ref{fig:preprocessing}. Consider input data \{$\mathcal{D}_1$, $\mathcal{D}_2$, ..., $\mathcal{D}_n$\} $\in \mathcal{D}$ is extracted from the raw data stored in the workload repository. 
The \textit{data extraction} filters relevant  attributes from raw data to improve the pattern learning and developing more intuitive correlations among extracted patterns. The \textit{data aggregation} operation is performed in which the extracted data is assembled as per the chosen prediction window-size (for example, five minutes) such as \{$\mathcal{D}_1$, $\mathcal{D}_2$, $\mathcal{D}_3$\} combines to produce an aggregated data sample $\mathcal{D}^{\ast}_1$. Similarly, \{$\mathcal{D}_{a}$, $\mathcal{D}_{a+1}$, $\mathcal{D}_{a+2}$\} and \{$\mathcal{D}_{n-2}$, $\mathcal{D}_{n-1}$, $\mathcal{D}_{n}$\}  aggregate to generate data samples $\mathcal{D}^{\ast}_a$ and $\mathcal{D}^{\ast}_n$, respectively.  The aggregated data samples are scaled in a specific range [$x_a$, $x_b$] and transformed into a normalized data samples \{$\mathcal{D}^{N}_1$, $\mathcal{D}^{N}_2$, ..., $\mathcal{D}^{N}_n$\} using Eq. (\ref{normalisation}), where  $\varpi^{In}_{min}$ and $\varpi^{In}_{max}$ are the minimum and maximum values of the input data set, respectively and the normalized data vector is denoted as $\hat{\varpi^{In}}$, which is a set of all normalized input data values.  The values of $x_a$ and $x_b$ were set equals to 0.0001 and 0.999, respectively for the experiments.  These normalized values  are organized into two dimensional input and output matrices denoted as $\varpi^{In}$ and $\varpi^{Out}$, respectively as stated in Eq. (\ref{eqn:2}):
 \begin{gather}
 \label{normalisation}
 \hat{\varpi^{In}}=x_a +\frac{ d_i- \varpi^{In}_{min}}{\varpi^{In}_{max}-\varpi^{In}_{min}}\times(x_b) \\
\label{eqn:2}
\resizebox{0.443\textwidth}{!}{$  
	\varpi^{In}= 
	\left[ {\begin{array}{cccc}
		\varpi_1 & \varpi_2 & .... & \varpi_z\\
		\varpi_2 & \varpi_3 & .... & \varpi_{z+1} \\
		.    &    .     & .... &    .     \\
		\varpi_m & \varpi_{m+1} & .... & \varpi_{z+m-1} \\    
		\end{array} } \right]
	\varpi^{Out}= 
	\left[ {\begin{array}{c}
		\varpi_{z+1} \\
		\varpi_{z+2}  \\
		.      \\		
		\varpi_{z+m}    
		\end{array} } \right] $}
 \end{gather} 

 Accordingly, prediction sliding window is prepared as shown in the block of sliding window in  Fig. \ref{fig:preprocessing}. These normalized data samples are divided into three categories namely, \textit{training data}, \textit{validation data}, and \textit{testing data}. A \textit{Machine Learning Algorithm} is appointed as a prediction model which receives training data to allow specific pattern learning during iterative learning or optimization  process. Thenafter, this  prediction model is evaluated using any error evaluation function such as RMSE or MSE, which is tested for accuracy. If the desired accuracy is achieved, a \textit{Trained Prediction Model} is obtained and validation data is passed into it again to check for prediction accuracy. If the optimal accuracy is achieved, a \textit{Validated Prediction Model} is established. Finally, the test/unseen data is passed to the validated prediction model and accuracy evaluation is performed. If consecutively, the desired accuracy is achieved, a \textit{Final Prediction Model} is deployed, otherwise, the respective stage of the prediction models reversed back to the iterative learning process. The performance is measured as predicted output is achieved for cloud resource management.
 
\section{{Research Methodology}}
{This section elaborates the review methodology in detail. The review procedure includes the gathering of major  cloud workload prediction papers wherein the proposed approaches are driven from machine learning algorithms and concepts. The pioneering quality prediction models, published in top-notch journals and conference databases such as IEEE, ACM Digital Library, Elsevier, Springer, Wiley are searched, studied and analysed for comparative study. Furthermore, the collected papers are refined by the identification of primary studies based on underlying proposed approaches, then application of a specific inclusion criteria for grouping the paper based on similar approaches or having overlapping features into a common class/category.  To avoid any biasness during research, the review process in the remaining sections is developed by one of the authors, and finalized by the other co-author via discussions, and iterative review methods. The existing prediction models are thoroughly explored by distinct authors to ensure the completeness of the proposed study and inter-performance comparison. While selecting related work corresponding to each category, the average and below average research work are filtered and avoided so as to present a clear and concise review of the best of the existing workload prediction approaches. Accordingly,  a classification and taxonomical representation is presented in Fig. \ref{fig:CloudComputing} which designates the existing prediction models into  specific classes and sub-classes of the supervised machine learning approaches based on their conceptual and operational characteristics.} 
\tikzset{
	process/.style={
		rectangle, 
		drop shadow,
		rounded corners,		
		font=\scriptsize\sffamily,
		minimum width=5.0cm, 
		minimum height=1cm, 
		align=center,
		draw=black, 
		fill=red!10! 
	},
	process1/.style={
		rectangle, 
		drop shadow,
		rounded corners,
		font=\scriptsize\sffamily,
		minimum width={width("Evolutiona")},
		minimum height=.7cm, 
		align=center, 
		draw=black, 
		fill=black!7
	},
	process2/.style={
		rectangle, 
		drop shadow,
		rounded corners,
		font=\scriptsize\sffamily,
		minimum width={width("LPAW Au")}, 
		minimum height=0.3cm, 
		align=center, 
		draw=black, 
		fill=green!7
	},
	process3/.style={
		rectangle,
		drop shadow,
		rounded corners,
		font=\scriptsize\sffamily,
		minimum width={width("Autoenco")}, 
		minimum height=0.7cm, 
		align=center, 
		draw=black, 
		fill=yellow!20
	},
	process4/.style={
		rectangle, 
		drop shadow,
		rounded corners,
		font=\scriptsize\sffamily,
		minimum width={width("LPAW Au")},
		minimum height=0.1cm, 
		align=center, 
		draw=black, 
		fill=cyan!15
	},
	process5/.style={
		rectangle, 
		drop shadow,
		rounded corners,
		font=\scriptsize\sffamily,
		minimum width={width("LPAW Aus")},
		minimum height=0.1cm, 
		align=center, 
		draw=black, 
		fill=orange!15
	},
	process6/.style={
		rectangle, 
		drop shadow,
		rounded corners,
		font=\scriptsize\sffamily,
		minimum width={width("LPAW A")}, 
		minimum height=0.1cm, 
		align=center, 
		draw=black, 
		fill=purple!14
	},    
	process7/.style={
		rectangle, 
		drop shadow,
		rounded corners,
		font=\scriptsize\sffamily,
		minimum width={width("LPAW A")},  
		minimum height=0.1cm, 
		align=center, 
		draw=black, 
		fill=teal!15
	}, 
	process8/.style={
		rectangle, 
		drop shadow,
		rounded corners,
		font=\scriptsize\sffamily,
		minimum width={width("LPAW A")},  
		minimum height=0.1cm, 
		align=center, 
		draw=black, 
		fill=olive!20
	},    
	process9/.style={
		rectangle, 
		drop shadow,
		rounded corners,
		font=\scriptsize\sffamily,
		minimum width={width("LPAW A")}, 
		minimum height=0.3cm, 
		align=center, 
		draw=black, 
		fill=violet!12
	},   
	process10/.style={
		rectangle, 
		drop shadow,
		rounded corners,
		font=\scriptsize\sffamily,
		minimum width={width("LPAW A")},  
		minimum height=0.1cm, 
		align=center, 
		draw=black, 
		fill=red!15
	},    
	arrow/.style={draw,thick,->,>=stealth},
	dec/.style={
		ellipse, 
		align=center, 
		draw=black, 
		fill=green!15
	},
}

\begin{center}
	\begin{figure*}[htbp!]
		\hspace{-2cm}
		\centering
		\resizebox{0.75\textwidth}{!}{	\begin{tikzpicture}[
			node distance=0.65cm,
			every edge/.style={arrow}
			]
			\node (c1)[process] at (50,0)  {Cloud Workload Prediction Models \\ Machine Learning Approaches};
			\node(c2) [process1, below left=of c1,text centered,yshift=-4pt,xshift=-110pt] { Evolutionary \\ Learning};
			\node (c3) [process1,right=of c2,text centered, xshift=80pt,yshift=-2pt] {Deep \\ Learning};
			\node(c4)[process1,right=of c3,text centered, xshift=90pt]{Hybrid \\ Learning};
			\node(c5)[process1,right=of c4,text centered, xshift=10pt]{Ensemble \\ Learning};
			\node(c6)[process1,right=of c5,text centered, xshift=10pt]{Quantum \\ Learning};   
			\node(c21)[process2,below=of c2,text centered,yshift=-2pt, xshift=10pt]{ANN+S\\ADE \cite{kumar2018workload}};  
			\node(c22)[process2,below=of c21,text centered,yshift=10pt,xshift=0pt]{ANN-BA\\DE \cite{kumar2020biphase}}; 
			\node(c23)[process2,below=of c22,text centered,yshift=10pt,xshift=0pt]{ANN-B\\HO \cite{kumar2021self}};
			\node(c24)[process2,below=of c23,text centered,yshift=10pt,xshift=0pt]{SDWF \\ \cite{kumar2021self}};
			\node(c25)[process2,below=of c24,text centered,yshift=10pt,xshift=0pt]{FLGAPS\\ONN  \cite{khorsand2018fahp}};
			\node(c31)[process3,right=of c21,text centered,xshift=-8pt]{LSTM};
			\node(c32)[process3,right=of c31,text centered,xshift=-8pt]{Autoencoder};
			\node(c33)[process3,right=of c32,text centered,xshift=8pt]{Deep Belief\\ Network};
			\node(c34)[process3,right=of c33,text centered,xshift=-10pt]{Deep Neural\\ Network};
			\node(c311)[process4,below=of c31,text centered,xshift=10pt,yshift=2pt]{LSTM \\ \cite{kumar2018long}};    
			\node(c312)[process4,below=of c311,text centered,xshift=0pt, yshift=10pt]{2D LSTM \\ \cite{tang2019large}};   
			\node(c313)[process4,below=of c312,text centered,xshift=0pt, yshift=10pt]{Bi-LSTM \\ \cite{gao2020task} };  
			\node(c314)[process4,below=of c313,text centered,xshift=0pt,yshift=10pt]{Crystal\\ ILP \cite{ruan2021workload}};  
			\node(c315)[process4,below=of c314,text centered,xshift=0pt,yshift=10pt]{FEMT-LSTM \\
				\cite{ruan2022cloud}};
			\node(c321)[process5,below=of c32,text centered,xshift=16pt, yshift=2pt]{Encoder+\\LSTM \cite{tuli2021start}};    
			\node(c322)[process5,below=of c321,text centered,xshift=0pt,yshift=10pt]{CP Autoen\\-coder \cite{zhang2018efficient}};   
			\node(c323)[process5,below=of c322,text centered,xshift=0pt,yshift=10pt]{LPAW Autoe\\-ncoder \cite{chen2019towards}};  
			\node(c324)[process5,below=of c323,text centered,xshift=0pt,yshift=10pt]{GRUED\\ \cite{peng2018multi}};
			\node(c331)[process6,below=of c33,text centered,xshift=10pt,yshift=2pt]{DBN+R\\BN \cite{qiu2016deep}};    
			\node(c332)[process6,below=of c331,text centered,xshift=0pt,yshift=10pt]{DBN+O\\ED \cite{zhang2017resource}};   
			\node(c333)[process6,below=of c332,text centered,xshift=0pt,yshift=10pt]{DP-CU\\PA \cite{wen2020cpu}};  
			\node(c341)[process7,below=of c34,text centered,xshift=10pt,yshift=2pt]{es-DNN\\ \cite{xu2022esdnn}};    
			\node(c342)[process7,below=of c341,text centered,xshift=0pt,yshift=10pt]{DNN+\\MVM \cite{bhagtya2021workload}};   
			\node(c343)[process7,below=of c342,text centered,xshift=0pt,yshift=10pt]{DNN-\\PPE \cite{li2016learning}}; 
			\node(c344)[process7,below=of c343,text centered,xshift=0pt,yshift=10pt]{SG-LSTM \\ \cite{bi2019deep}}; 
			\node(c41)[process8,below=of c4,text centered,xshift=16pt,yshift=0pt]{ADRL \\\cite{kardani2020adrl}};
			\node(c42)[process8,below=of c41,text centered,xshift=0pt,yshift=10pt]{Bi-Hyp\\-rec \cite{karim2021bhyprec}};
			\node(c43)[process8,below=of c42,text centered,xshift=0pt,yshift=10pt]{BG-LSTM \\\cite{bi2021integrated}};
			\node(c44)[process8,below=of c43,text centered,xshift=0pt,yshift=10pt]{HPF-DNN \\ \cite{chen2015self}};
			\node(c45)[process8,below=of c44,text centered,xshift=0pt,yshift=10pt]{FAHP\\ \cite{khorsand2018fahp}};
			\node(c46)[process8,below=of c45,text centered,xshift=0pt,yshift=10pt]{ACPS\\\cite{liu2017adaptive}};
			\node(c47)[process8,below=of c46,text centered,xshift=0pt,yshift=10pt]{LSRU \\ \cite{shuvo2020lsru}};    
			\node(c51)[process9,below=of c5,text centered,xshift=13pt]{KSE+WMC\\ \cite{singh2014ensemble}};
			\node(c52)[process9,below=of c51,text centered,xshift=0pt,yshift=10pt]{FAST \\ \cite{feng2022fast}};
			\node(c53)[process9,below=of c52,text centered,xshift=0pt,yshift=10pt]{SGW-S \\ \cite{bi2019temporal}};
			\node(c54)[process9,below=of c53,text centered,xshift=0pt,yshift=10pt]{ClIn \\ \cite{kim2020forecasting}};
			\node(c55)[process9,below=of c54,text centered,xshift=0pt,yshift=10pt]{AMS \\ \cite{iqbal2019adaptive}};
			\node(c56)[process9,below=of c55,text centered,xshift=0pt,yshift=10pt]{E-ELM\\ \cite{kumar2020ensemble}};
			\node(c57)[process9,below=of c56,text centered,xshift=0pt,yshift=10pt]{SF-Cluster\\ \cite{chen2015self}};
			\node(c61)[process10,below=of c6,text centered,xshift=13pt,yshift=0pt]{EQNN\\ \cite{singh2021quantum}};
			
			\draw[|-,-|,->, thick,] (c1.south) |-+(0,-1em)-| (c2.north);
			\draw[|-,-|,->, thick,] (c1.south) |-+(0,-1em)-| (c3.north);
			\draw[|-,-|,->, thick,] (c1.south) |-+(0,-1em)-| (c4.north);
			\draw[|-,-|,->, thick,] (c1.south) |-+(0,-1em)-| (c5.north); 
			\draw[|-,-|,->, thick,] (c1.south) |-+(0,-1em)-| (c6.north);        
			\foreach \value in {1,...,5}
			\draw[->] (c2.210) |- (c2\value.west);
			\draw[|-,-|,->, thick,] (c3.south) |-+(0,-1em)-| (c31.north);
			\draw[|-,-|,->, thick,] (c3.south) |-+(0,-1em)-| (c32.north);
			\draw[|-,-|,->, thick,] (c3.south) |-+(0,-1em)-| (c33.north);
			\draw[|-,-|,->, thick,] (c3.south) |-+(0,-1em)-| (c34.north);
			\foreach \value in {1,...,5}
			\draw[->] (c31.210) |- (c31\value.west);
			\foreach \value in {1,...,4}
			\draw[->] (c32.220) |- (c32\value.west);   
			\foreach \value in {1,...,3}
			\draw[->] (c33.212) |- (c33\value.west);    
			\foreach \value in {1,...,4}
			\draw[->] (c34.210) |- (c34\value.west);  
			\foreach \value in {1,...,7}
			\draw[->] (c4.220) |- (c4\value.west);  
			\foreach \value in {1,...,7}
			\draw[->] (c5.215) |- (c5\value.west);     
			\foreach \value in {1}
			\draw[->] (c6.215) |- (c6\value.west);  
			\end{tikzpicture}}

		\caption{Classification and Taxonomy of Machine learning based Workload Prediction Models}
		\label{fig:CloudComputing}
		
	\end{figure*}
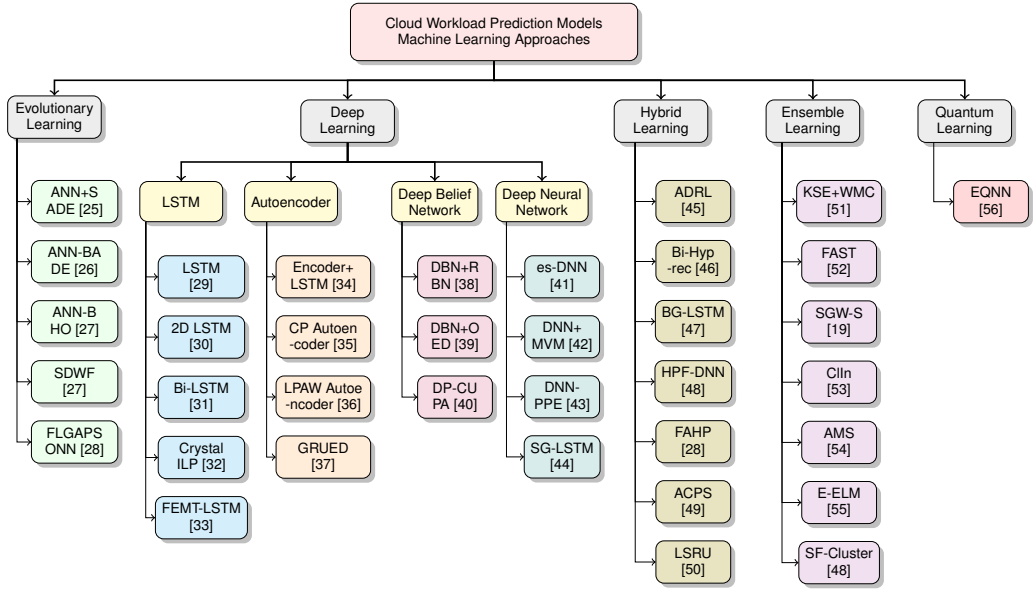
\end{center}
Correspondingly, the five exclusive workload prediction classes are designated as \textit{Evolutionary Learning}, \textit{Deep Learning}, \textit{Hybrid Learning}, \textit{Ensemble  Learning}, and \textit{Quantum Learning}. {In Evolutionary learning class, the candidate approaches: ANN+SADE \cite{kumar2018workload}, ANN-BADE \cite{kumar2020biphase},   ANN-BHO \cite{kumar2021self}, SDWF \cite{kumar2021self}, and FLGAPSONN  \cite{khorsand2018fahp} have applied evolutionary optimization alogorithms for the learning process or weight update process of neural network layers. The ample of works subject to Deep learning are further differentiated into four sub-classes including \textit{Long Short-Term Memory} (LSTM) cell, \textit{Autoencoder}, \textit{Deep Belief Network}, and \textit{Deep Neural Network}. The sub-class LSTM includes LSTM \cite{kumar2018long}, 2D LSTM \cite{tang2019large}, Bi-LSTM \cite{gao2020task}, Crystal ILP \cite{ruan2021workload}, and FEMT-LSTM 
\cite{ruan2022cloud} which are pre-dominantly based on functionality of LSTM models. Autoencoder sub-category consists of Encoder+LSTM \cite{tuli2021start}, CP Autoencoder \cite{zhang2018efficient}, LPAW Autoencoder \cite{chen2019towards}, and GRUED \cite{peng2018multi}  are derived by applying some useful modification in the traditional autoencoders. Similarly, DBN+RBN \cite{qiu2016deep}, DBN+OED \cite{zhang2017resource}, DP-CUPA \cite{wen2020cpu}; and es-DNN \cite{xu2022esdnn}, DNN+MVM \cite{bhagtya2021workload}, DNN-PPE \cite{li2016learning}, SG-LSTM \cite{bi2019deep} are located with sub-class Deep Belief Network and sub-class Deep Neural Network, respectively. The  Hybrid learning class represents integration of several machine learning algorithms and methods which encompasses ADRL \cite{kardani2020adrl}, Bi-Hyprec \cite{karim2021bhyprec}, BG-LSTM \cite{bi2021integrated}, HPF-DNN  \cite{chen2015self}, FAHP \cite{khorsand2018fahp}, ACPS \cite{liu2017adaptive}, and LSRU  \cite{shuvo2020lsru}. Likewise, Ensemble learning involves  concept of base-learners and decision making to estimate the final outcome. KSE+WMC \cite{singh2014ensemble}, FAST \cite{feng2022fast}, SGW-S  \cite{bi2019temporal}, ClIn  \cite{kim2020forecasting}, AMS  \cite{iqbal2019adaptive}, E-ELM \cite{kumar2020ensemble}, and SF-Cluster \cite{chen2015self}     works enfold  Ensemble learning. Finally, the Quantum learning class comprises EQNN model \cite{singh2021quantum} which is developed using the principles of quantum computing and neural network learning.}  
 
 \section{Evolutionary Neural Network based Prediction Models}
\begin{figure}[!htbp]
 	\centering
 	\includegraphics[width=0.7\linewidth]{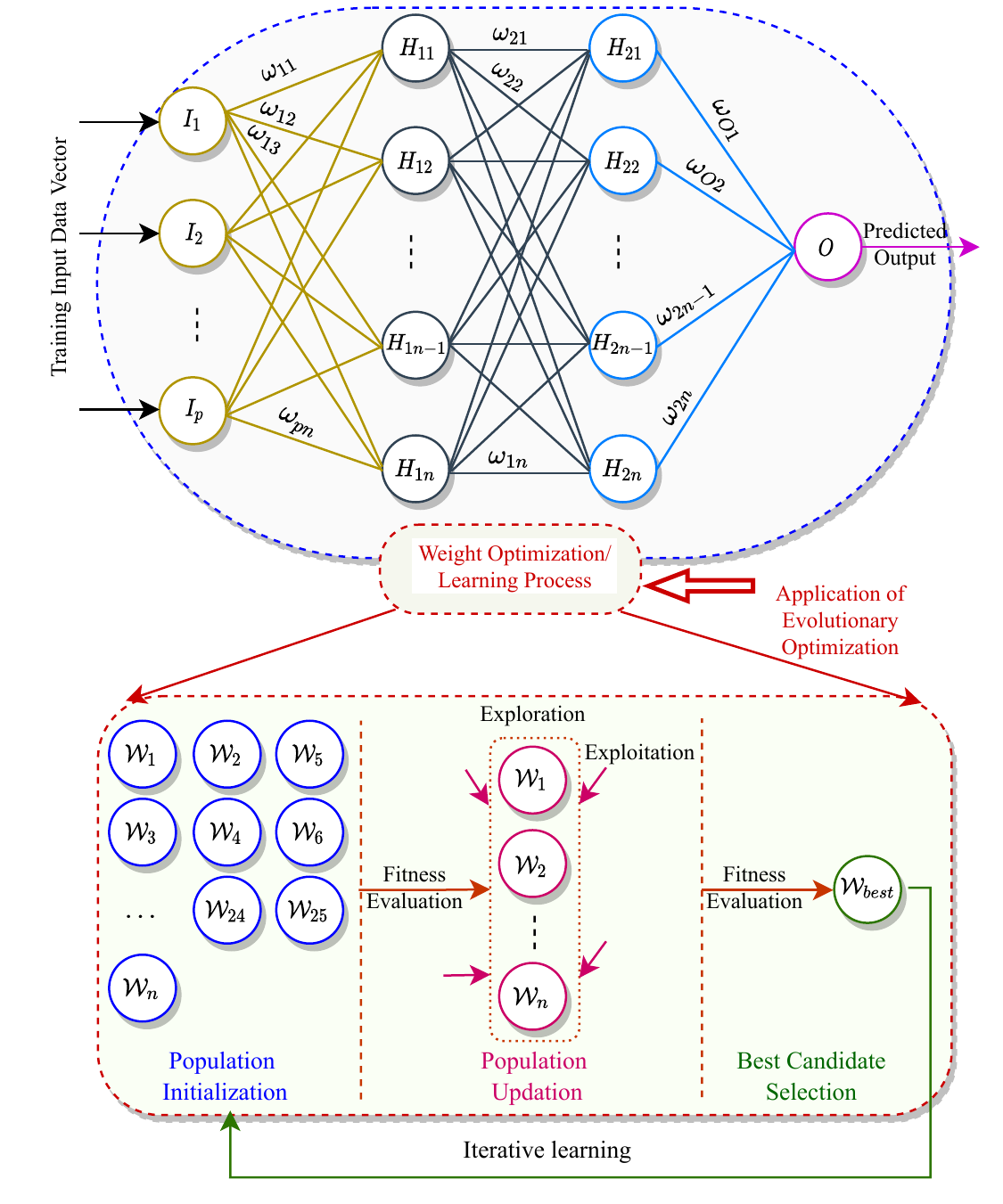}
 	\caption{ ENN based load prediction }
 	\label{fig:enn}
 \end{figure}

 The neural networks in which the learning process is achieved with the help of an evolutionary optimization algorithm are designated as \textit{Evolutionary Neural Networks} (ENN)s. Fig. \ref{fig:enn} depicts a schematic and operational view of an ENN. Consider a feed-forward neural network $n$-$p_1$-$p_2$-$q$ comprising of $n$, $q$ nodes in input and output layers, and $p_1$ and $p_2$ nodes in three consecutive hidden layers.  A training input data vector:  \{$\mathcal{D}_1$, $\mathcal{D}_2$, ..., $\mathcal{D}_n$\} $\in \mathcal{D}$ is passed to the input layer of this neural network, wherein each node of one layer is connected to all the nodes of the consecutive layer  with the help of synaptic/neural weights \{$w_1$, $w_2$, ..., $w_z$\} $\in \mathcal{W}$, $z$ is the total number of weight connections between any two consecutive neural layers. The forward propagation of training input vector ($\mathcal{D}$) is carried out via weighted connections  using Eqs. (\ref{l1}), (\ref{l2}), and (\ref{l3}), $\uplus$ is a linear function computed at each neuron; and $\mathcal{B}$ is  bias vector. The values of weight connections \{$w_1$, $w_2$, ..., $w_z$\}  determine impact of the input vector on the output vector of the neurons and decide strength of synaptic inter-connections between neurons.  
 \begin{gather}
 \uplus = (\mathcal{D}_1 \times w_1) + (\mathcal{D}_2 \times w_2) + ... +(\mathcal{D}_n \times w_n) \label{l1}\\
 \mathcal{D}^{\dagger}\cdot{\mathcal{W}^\dagger}=(\mathcal{D}_1 \times w_1) + (\mathcal{D}_2 \times w_2) + ... +(\mathcal{D}_n \times w_n) \label{l2}\\
 \uplus=	\mathcal{D}^{\dagger}\cdot{\mathcal{W}^\dagger} + \mathcal{B} \label{l3}
 \end{gather}  
 The ENN prediction model learns by adjusting the values of bias and inter-connection weights \{$w_1$, $w_2$, ..., $w_n$\} $\in \mathcal{W}^\dagger$ of ENN with the aim of minimizing the prediction error. This learning process is achieved by applying different evolutionary optimization algorithm which selects the most optimal network from the random population of $\mathcal{Z}$  networks \{$w^\dagger_1$, $w^\dagger_2$, ..., $w^\dagger_{\mathcal{Z}}$\}.  The algorithm repeatedly optimizes the values of neural weight connections by updating the population of networks exploring and exploiting  the diverse population of networks extensively. During successive epoch, the best  network candidate is chosen by evaluating a fitness function such as prediction error estimation using Root Mean Squared Error (RMSE), Mean Absolute Percentage Error (MAPE) etc. 
 To allow the online optimization of ENNs, the learning process is periodically repeated  with the previous and most recent samples of workloads  for their training and re-training  in an off-line mode  while analysing the live workloads concurrently in the real-time. There are several approaches proposed using ENN for cloud workload prediction and this section further aimed at providing comprehensive discussion and analysis of these approaches. 
 
 \par 
 An ENN-based cloud workload prediction approach is proposed in \cite{kumar2018workload} wherein a three layered feed-forward neural network is trained using a Self adaptive Differential Evolution (SaDE). During the learning process, the population of networks is updated by applying exploration and exploitation operations using three mutation strategies selectively followed by uniform crossover. This approach produced improved accuracy over Backpropagation trained neural network (BPNN) \cite{prevost2011prediction} because of multidimensional learning in former as compared with optimizing single solution in later approach. Kumar et al. \cite{kumar2020biphase} have proposed a Biphase adaptive Differential Evolution (BaDE) learning based neural network 
 that adopted a dual adaptation viz., at level of crossover during exploitation
 process and mutation in exploration phase to improve the learning efficiency of neural network. As a consequence, this work outperformed SaDE \cite{kumar2018workload} in terms of prediction accuracy. An auto-adaptive neural network is developed in \cite{saxena2020auto} wherein the network connection weights are adjusted with Tri-adaptive Differential Evolution  (TaDE) algorithm. In this approach, the 
 adaptation is appointed at level of crossover, mutation, and control parameters generation level which allows enhanced learning the prediction model. Kumar et al. \cite{kumar2016dynamic} have used a BlackHole Optimization (BHO) algorithm to optimize neural weights and develop a workload prediction model for dynamic resource scaling. This evolutionary optimization algorithm updates the movement of the stars i.e., randomly intialized network vectors and track their position whether reaching an event horizon. Further, this work was enhanced by modifying the existing BHO algorithm as enhanced BHO (i.e., E-BHO) in \cite{kumar2021self} by including concepts of local and global blackhole (i.e., best solution) during iterative learning process of the feed-forward neural network. Also, this approach has  computed the deviation in recent forecasts and applied it to enhance the accuracy of the forthcoming predictions.
 Malik et al. \cite{khorsand2018fahp} have proposed an ENN based multi-resource utilization prediction approach. In this work, Functional Link Neural Network (FLNN) with a hybrid evolutionary algorithm comprising of Genetic Algorigm (GA) and Particle Swarm Optimization (PSO) is applied for neural network weight adjustment during learning process. It has been compared with FLNN,  FLNN with GA (FLGANN), and FLNN with PSO (FLPSONN) and validated its performance against these methods.
 
 \par A pandect summary of existing ENN-based workload prediction models are provided in Table \ref{summaryNN} which highlights the major features, implementation details, performance metrics, parameter tunings and intended computational complexities during the learning process. The approaches discussed in \cite{kumar2018workload, kumar2020biphase, saxena2020auto} are based on Differential Evolution which needs tuning of multiple control parameters including crossover-rate, mutation-rate, keeping track of dynamic or fixed learning-period, maintaining records of the number of failure and successful candidates during each epoch. While the prediction approaches entailed in \cite{kumar2016dynamic,  kumar2021self} employing BHO keeps track of event-horizon radius only and updates the next generation population depending on the comparison of the fitness value of each candidate with the horizon radius. Hence, it can be analysed that  the consumption of training time and involved computational complexity, number of training epochs  are higher for DE based prediction approaches as compared with that of BHO based prediction. Based on the aforementioned factors, \cite{kumar2021self} is the most admissible among all the discussed approaches.
 
 \begin{table*}[htbp]	
 	\caption{{ Summary and Comparison of ENN-based Prediction Approaches}}
 	\label{summaryNN}
 	\centering
 	\resizebox{0.99\textwidth}{!}{
 		\begin{tabular}{p{2.0cm} p{1.5cm}  p{6.5cm} p{2.5cm}p{2.0cm}  p{1.8cm} p{1.1cm} p{3.6cm}} \hline
 			{\textbf{Notable Contributors (Timeline)} }&	\textbf{Model/ Approach/ Framework}&{\textbf{Workflow/ Strategy}}&\textbf{Datasets}&\textbf{Implementation/ Simulation tool}&{\textbf{Predicted parameters}}&{\textbf{Error metrics}}&{\textbf{Results or Remarks}} \\ \hline \hline
 			
 			Kumar et al. \cite{kumar2018workload} (2018)	&{ANN-SaDE} & Feed-forward neural network is trained with self-adaptive differential evolutionary algorithm       &{NASA and Saskatchewan}  &  MATLAB&Number of requests per unit time& MSE & reduced error up to 0.001 and accuracy improved by 168 times over BP   \\ \hline
 			Kumar et al.		 \cite{kumar2020biphase} (2020)&{ANN-BaDE} & three mutations stategies and three crossover strategies based adaptation within DE algorithm optimizes ANN  &{NASA, Saskatchewan, Google Cluster traces }& Python&Number of requests, CPU, memory usage  &{MSE}& accuracy improved up to  up to 91\% and 97\% over SaDE and
 			BP, respectively \\ \hline
 			Saxena et al. \cite{saxena2020auto} (2020)		& {ANN-TaDE} &three mutations stategies and three crossover strategies with control parameter-tuning based adaptations within DE algorithm optimizes ANN   &{NASA and Saskatchewan}& Python& Number of requests per unit time  &{RMSE}& accuracy improved by 97.4\% and 94.8\% over BP and SaDE, respectively \\ \hline
 			Kumar et al. \cite{kumar2016dynamic} (2016) & Neural Network with BlackHole Optimization& Blackhole algorithm is utilized in the learning process of neural network provoded with pre-processed training data    & HTTP traces from NASA, Calgary and Saskatchewan web
 			servers  &MATLAB  & Number of requests per unit time & {MSE}& error reduced upto 134 times over BP  \\ \hline
 			Kumar et al.		{ \cite{kumar2021self}}  (2021)& Self directed workload forecasting method (SDWF)&   
 			the forecasting error trend is captured by computing the deviation in recent forecasts and applied 	to enhance the accuracy of further predictions &  {NASA and Saskatchewan} HTTP traces, Google Cluster & MATLAB &Number of requests, CPU, memory usage   &{MSE}& error up to 99.99\% over
 			compared methods  \\ \hline
 			Malik et al. \cite{khorsand2018fahp} (2022) & FLGAPSONN: FLNN +GA+ PSO &  Pre-processed training data is passed to FLNN which is optimized with hybrid algorithm of GA and PSO &  Google Cluster traces & Python &CPU, memory, disk usage& MAE & 	 improved accuracy by 21.87\%, 13.75\%, and 30.55\% over FLPSONN, FLGANN, and FLNN, respectively \\ \hline
 			
 		\end{tabular}
 	}
 \end{table*}
 
\section{Deep Learning based Prediction Models}
\begin{figure}[!htbp]
	\centering
	\includegraphics[width=0.8\linewidth]{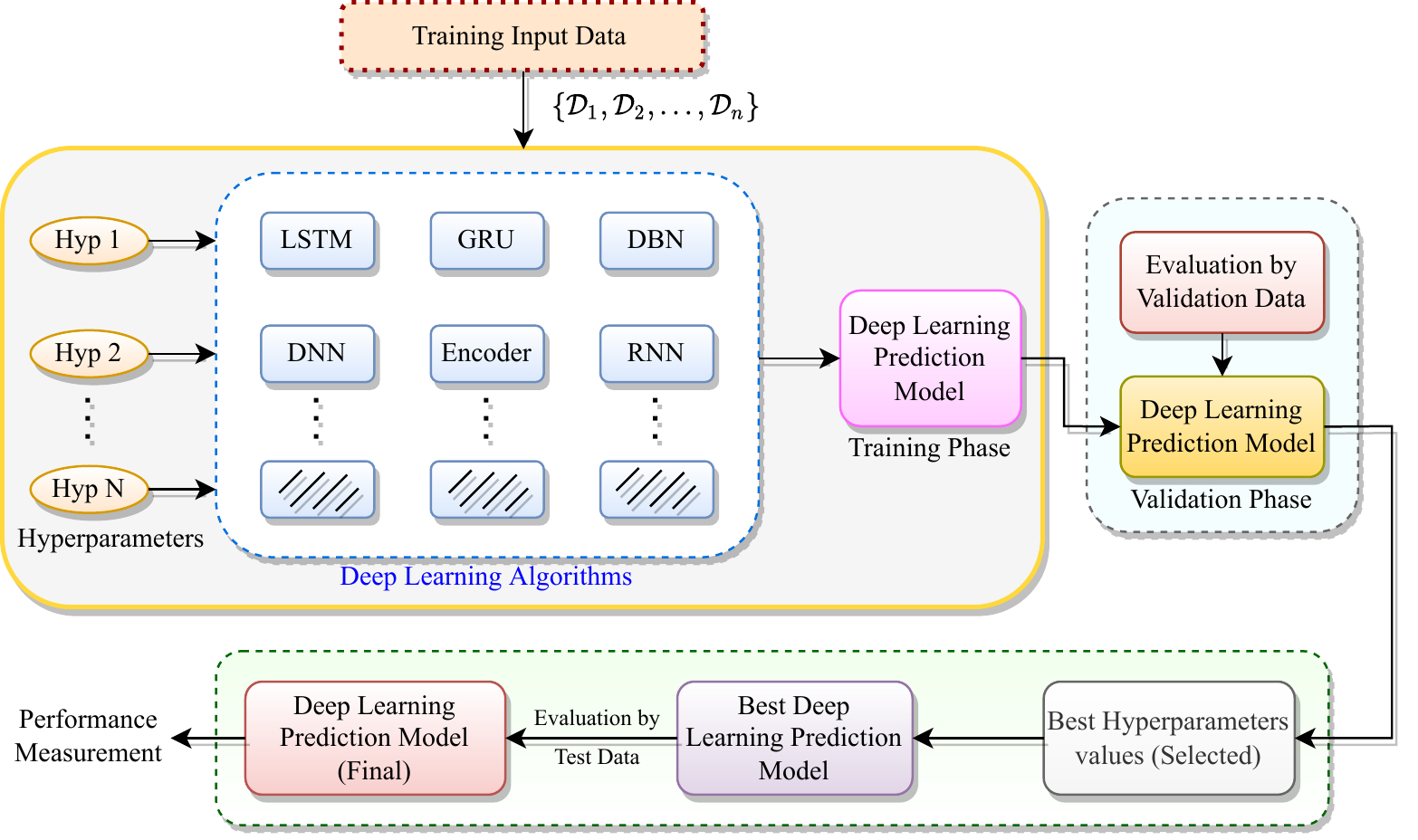}
	\caption{Deep Learning based Prediction Operative View}
	\label{fig:deep-learning}
\end{figure}

Deep Learning  models are a class of prediction models which have immensely influenced the field of cloud computing. A conceptual and operational design of deep learning  strategy is illustrated in Fig. \ref{fig:deep-learning}, wherein, the pre-processed training input data samples: \{$\mathcal{D}_1$, $\mathcal{D}_2$, ..., $\mathcal{D}_n$\} are passed into a deep learning algorithms such as Long Short-Term   Memory (LSTM), Gated Recurrent Units (GRU), Deep Belief Networks (DBN), Deep Feed-forward Neural Network (DNN), Autoencoders, Recursive Neural Network (RNN) and so on for the learning process. The essential hyperparameters \{$Hyp 1$, $Hyp 2$, ..., $Hyp N$\} of the respective deep learning algorithms are tuned  and re-tuned periodically to create and update the deep learning based prediction model. The trained model is further evaluated using validation data to estimate its performance and accuracy. Accordingly, a deep learning based model optimized with best or most admissible hyperparameters is obtained. 
For instance, LSTM based deep learning algorithm is applied for the prediction. The resource usage of actual load ($\mathcal{Z}^{Ac}_{\mathcal{RU}}$) stored as a historical workload is fed as input into neural network input layer for prediction of future resource usage. LSTM-RNN based prediction model comprises of cells containing four neural network  layers, where previous cell ($\mathcal{X}^{t-1}$) information is passed to current cell ($\mathcal{X}^{t}$).  The first layer ($\mathcal{G}_1$) applies Eq. (\ref{lstm1}) to decide the amount of previous resource usage information ($\mathcal{CF}_{\mathcal{RU}_i}^t$) which is tranferred to the next state; where $\mathcal{WT}$ is weight matrix, $\mathcal{B}$ is a bias value, $\mathcal{Z}^{Pr}_{t-1}$ and $\mathcal{Z}^{Pr}_t$ are previous output and current input, respectively. The cell state is updated using two network layers viz., $sigmoid$ layer ($\mathcal{G}_2$) which decides the values to be updated ($\mathcal{I}^t$) using Eq. (\ref{lstm2}), and $tanh$ layer for generation of a new candidate values vector ($\hat{\mathcal{X}}^t$) using Eq. (\ref{lstm3}). Finally, Eq. (\ref{lstm4}) combines both outputs to update cell state.   
\begin{equation}\label{lstm1}
\mathcal{CF}_{\mathcal{RU}_i}^t = \mathcal{G}_1(\mathcal{WT}_{\mathcal{CF}}\cdotp [\mathcal{Z}^{Pr}_{t-1}, \mathcal{Z}^{Ac}_t] +\mathcal{B}_{\mathcal{RU}})
\end{equation}
\begin{equation} \label{lstm2}
\mathcal{I}^t = \mathcal{G}_2(\mathcal{WT}_{\mathcal{I}}\cdot [\mathcal{Z}^{Pr}_{t-1}, \mathcal{Z}^{Ac}_t] +\mathcal{B}_{\mathcal{I}})
\end{equation}
\begin{equation} \label{lstm3}
\hat{\mathcal{X}}^t = {tanh}(\mathcal{WT}_{\mathcal{X}}\cdot [\mathcal{Z}^{Pr}_{t-1}, \mathcal{Z}^{Ac}_t] +\mathcal{B}_{\mathcal{X}})
\end{equation}
\begin{equation} \label{lstm4}
\mathcal{X}^t =\mathcal{CF}_{\mathcal{RU}_i}^t \times \mathcal{X}^t  + \hat{\mathcal{X}}^{t-1} \times \mathcal{I}^t 
\end{equation}
The RU of predicted traffic ($\mathcal{Z}^{Pr}_{t+1}$) for different VMs hosted on a server are aggregated to determine any overload proactively and alleviate it by migrating VMs with highest predicted RU to an efficient server. A comprehensive survey of the existing workload prediction models belonging to four distinct categories of deep learning approaches are discussed in the subsequent sections.
\subsection{LSTM-RNN}
  A fine-grained cloud workload prediction model using long short-term memory based recurrent neural network (LSTM-RNN) is presented in \cite{kumar2018long} which is capable of learning a long-term dependencies and producing a high accuracy for host load prediction. Tang \cite{tang2019large} has proposed a two-dimensional LSTM neural network cell structure by utilizing a hidden layer week-based dependence and weights parallelization algorithm. This work has improved LSTM algorithm by providing the mathematical description of parallel LSTM algorithm and its optimization with an error back propagation method. Its performance is validated using the real workload of the  Shanghai Supercomputer Center. Tuli et al. \cite{tuli2021start} have  proposed an automatic straggler (slow processing tasks) prediction and mitigation method for cloud environment using an encoder LSTM network that addressed heterogeneous host and volatile task characteristics. The encoder analyses the resource usage and load information and passes the information to the LSTM. Further, an exponential moving average of input matrices is taken into account to prevent the LSTM model from diverging. A  storage workload prediction approach named CrystalLP based on LSTM
neural network is introduced in \cite{ruan2021workload}. In this approach, a storage workload time-series model is developed which collects the intended workload patterns that helps in  precise and adaptive scheduling with load balancing. Thereafter, LSTM based workload predictor is implemented which is trained or optimized with an algorithm composed of an integration of stochastic gradient descent (SGD) together with the Adam optimizer. Gao et al. \cite{gao2020task} have presented a multi-layer Bi-directional Long Short Term Memory (Bi-LSTM)  based task failure prediction algorithm. It comprises of one input layer, two Bi-LSTM layers, one output layer and the Logistic Regression (LR) layer to predict whether the tasks are failed or finished. Unlike traditional LSTM which uses only forward state, Bi-LSTM operates on both forward and backward states to allow more accurate estimation of the weights of both closer and farther input features. Ruan et al. \cite{ruan2022cloud} have established a turning point prediction model for cloud server workload forecasting considering cloud
workload features. Thenafter, a  cloud feature-enhanced deep learning model with rule-filtering
based Piecewise Linear Representation (PLR) algorithm is build for workload turning point prediction. The performance evaluation of this model illustrated its prediction accuracy effectiveness in terms of improvement in  F1 score over existing state-of-the-art methods.

\subsection{Auto-encoder}
An efficient canonical polyadic (CP) decomposition based deep learning model  is proposed in \cite{zhang2018efficient} for prediction of industrial workloads in cloud, wherein, a CP auto-encoder is constructed by converting a basic autoencoder  into tensor space with the help of bijection.  In this model,   the basic auto-encoder in the CP decomposition format is  followed by the stacked autoencoder model in the CP decomposition format. The stacked
autoencoder is created to learn the relevant features of the workload information  and the CP decomposition is employed to compress the features substantially
for improving the training efficiency. Chen et al. \cite{chen2019towards} have established a  deep 
Learning based Prediction Algorithm for cloud Workloads (L-PAW) which included 
a Top-Sparse Auto-encoder (TSA) for an effective extraction of 
the essential representations of workloads. This approach integrated GRU and recurrent neural network (RNN) to evict the long-term memory dependencies for prediction of forthcoming cloud workloads with enhanced accuracy. 
\subsection{Deep Belief Network}
Qiu et al. \cite{qiu2016deep} have proposed a Deep Belief Network (DBN) composed of multiple-layered Restricted Boltzmann Machines (RBMs) and a regression layer for prediction of cloud workloads. In this model, DBN extracts the high level features from all VMs and the regression layer is used to predict the forthcoming load on VMs. It learns significant patterns efficiently using prior knowledge in an unsupervised manner.
Zhang et al. \cite{zhang2017resource} developed a DBN approach based load prediction model which is a stacked RBM and used Backpropagation algorithm to minimize its loss function. It incorporated analysis of variance and Orthogonal Experimental Design (OED) techniques into the parameter learning of DBN and have achieved a high prediction accuracy over ARIMA. Also, a similar DBN-based approach  is presented in \cite{zhang2015towards} which can capture high variances in cloud metric data without handcrafting specified feature for short term resource demands and long-term load prediction. 
Wen et al. \cite{wen2020cpu} have presented a DBN and Particle Swarm Optimization (PSO) based CPU usage prediction algorithm named as DP-CUPA. This algorithm includes three main steps: pre-processing of training data samples; adoption of autoregressive and grey models as base prediction models; and training of DBN. The PSO is utilized for estimation of  DBN parameters during learning process.

\subsection{Deep Neural Network}
 Xu et al. have proposed an efficient supervised learning-based Deep Neural Network (esDNN) algorithm \cite{xu2022esdnn} to extract and learn the features of historical data and accurately predict future workloads. The multivariate data is converted into supervised learning time series and a revised GRU is applied which can adapt   to the variances of workloads  to achieve accurate prediction and overcome the limitations of gradient  disappearance and explosion. 
 A DNN based workload prediction method (designated as DNN-MVM) is developed in \cite{bhagtya2021workload} to handle the workload prediction from multiple virtual machines. It employed a pre-processing and feature selection engine to handle data directly  from these virtual machines. The model classifies data based on historical loads to provide enhanced information or knowledge to the cloud  service provider for resource management and optimization. It is useful to predict the peak demands of resources in the future. This model is validated using Grid Workload Archive (GWA) dataset. Bi et al. \cite{bi2019deep} have proposed another DNN based workload prediction model wherein a logarithmic operation is performed ahead of task smoothening to minimize the standard deviation. Thenafter, a Savitzky-Golay (S-G) filter is
 applied to eliminate the extreme points and noise interference in the sequence of the  original data. The DNN based LSTM (SG-LSTM) is employed to extract complicated characteristics of a task time series. A Back Propagation Through Time (BPTT) algorithm is implemented by adopting  gradient clipping method  to eliminate a gradient exploding problem and optimization of the model is accomplished using Adam Optimizer. The model is evaluated with Google  Cluster dataset to confirm its efficacy over existing methods. A DNN learning based Power Prediction Engine  (DNN-PPE) is proposed in \cite{li2016learning} which includes data acquisition, data pre-treatment, and prediction modules. Recursive auto encoder is utilized for short-time fine-grained prediction that 
 can track the rapid changes of the power consumption within a
 data centre.

\begin{table*}[htbp]
\caption{{Comparative Summary of Deep Learning-based Prediction Approaches}}
\label{summaryDNN1}
\centering
\resizebox{0.99\textwidth}{!}{
	\begin{tabular}{p{2.0cm} p{1.5cm}  p{6.3cm} p{2.5cm}p{2.04cm}  p{1.8cm} p{1.11cm} p{3.6cm}} \hline
		{\textbf{Notable Contributors (Timeline)} }&	\textbf{Model/ Approach/ Framework}&{\textbf{Workflow/ Strategy}}&\textbf{Datasets}&\textbf{Implementation/ Simulation tool}&{\textbf{Predicted parameters}}&{\textbf{Error metrics}}&{\textbf{Results or Remarks}} \\ \hline \hline
		Kumar et al. \cite{kumar2018long} (2018)&	LSTM-RNN  & Combinations of networks in loop  retains learning information, performs specific operation to produce output for next network in loop & Web server HTTP traces&MATLAB   & No. of requests per unit time& MSE & reduced RMSE up to 0.00317 \\ \hline 
		Tang \cite{tang2019large} (2019)&	2-D LSTM  & Input data is analysed on week dependence basis and weights parallelization algorithm is used for improved optimization of LSTM & l workload of the Shanghai Supercomputer& Not mentioned & Workload over days & MSE &  accuracy improved for large-scale real-time cloud services\\ \hline
			
	Tuli et al. \cite{tuli2021start} (2021)	&	Encoder LSTM  & Encoder is used for input matrices preparation and LSTM predicts the load information & PlanetLab traces  & CloudSim and Python  & CPU, Memory, Bandwidth & MSE, MAPE &  reduced execution time, resource contention, energy and SLA violations by 13\%, 11\%, 16\% and 19\%\\ \hline
	Gao et al. \cite{gao2020task} (2020) &Bi-LSTM   & Training data propagates via input layer, two Bi-LSTM layers, one output layer and LR layer during learning process & 55,55,55 tasks  traces& Tensorflow in Python  &  task failure rate & F1-Score& 93\% accuracy and 87\% task failure correctly predicted\\ \hline
	Ruan et al.	\cite{ruan2021workload}	(2021)	&CrystalLP & Time-series model collects load patterns and LSTM trained with SGD+Adam optimizer predicts load& Web search archive SPC
	traces  & Keras library,  Python  & Request size & MAPE, RMSE, MAE & achieved 1.10\% improvement in MAPE, and  better performance in MAE over existing methods \\ \hline
	
	Ruan et al.	\cite{ruan2022cloud} (2022) & FEMT-LSTM &  a turning point prediction model considering cloud workload features followed by feature-enhanced deep learning model is developed  &  Google Cluster, Alibaba, HPC Grid workloads & Keras library,  Python & CPU usage & binary cross-entropy, F1, precision, recall& F1 score is improved by 6.6\% over existing approaches \\ \hline					
Zhang et al.	\cite{zhang2018efficient}	(2018)	&	CP auto-encoder  & canonical polyadic decomposition compresses the features and stacked auto-encoder learns the patterns for prediction &PlanetLab traces & MATLAB &  CPU utilization & MAPE, RMSE & achieves a higher training efficiency and
 prediction accuracy  for industrial workloads
 \\ \hline
			
Chen et al. \cite{chen2019towards} (2019)&L-PAW auto-encoder & TSA extracts workload patterns and GRU+RNN prdicts the upcoming load &TensorFlow 1.4.0, Python & DUX-cluster, Alibaba, and Google cluster &CPU, memory, disk I/O usage  &MSE, CDF & outperformed the accuracy of LSTM, RNN, GRU\\ \hline		 
			 
	Qiu et al. \cite{qiu2016deep} (2016)& DBN+RBMs  & DBN extracts significant patterns while learning and regression is used for prediction& PlanetLab traces  & CloudSim  & CPU utilization & MAPE &improved the performance up to 1.3\% over existing method \\ \hline 
Zhang et al.	\cite{zhang2017resource} (2017)	&	DBN+OED  & Pre-processed data is passed into DBN+OED model which is tuned with Backpropagation algorithm  & Google Cluster traces   & Python   &CPU, RAM usage &  MSE& MSE achieved in the range  [$10^{-4}$, $10^{-3}$] \\ \hline
Wen et al. \cite{wen2020cpu} (2020) & DP-CUPA  & DBN predicts the load information and PSO estimates the fitness values of its tuning parameters &Google Cluster traces  & not mentioned  & CPU usage & MSE, MAPE, MAE& outperformed autoregressive, DBN, Grey model \\ \hline
	Xu et al.	\cite{xu2022esdnn} (2022) &	es-DNN  & supervised learning converts multi-variate data into time-series and modified GRU is applied &  Alibaba and Google Cluster traces  &TensorFlow 2.2.0 in Python &  CPU usage per time-unit interval & MAPE, MSE, RMSE & reduced  number of active hosts efficiently and optimized cost  \\ \hline			
Bhagtya et al.	\cite{bhagtya2021workload} (2021)&	DNN-MVM &Selected data from multiple VMs is classified, preprocessed and passed to DNN-MVM for learning process & Grid Workload Archive (GWA) traces   &   Google Colab using Keras 	& CPU, Memory, and
			Disk Utilization & MSE &achieved more than 85\% prediction accuracy for each resource \\ \hline
	Bi et al. \cite{bi2019deep} (2019) &	SG-LSTM   & S-G filter provides smoothen data to LSTM for more accurate prediction &Google traces    & Python   & CPU, Memory &  Logarithmic RMSE, $R^2$ & outperformed BPNN, LSTM, SG-LSTM, and SG-BPNN 	\\ \hline
		
Li et al.	\cite{li2016learning} (2016)&	DNN-PPE & Preprocessed data is passed to Recursive Auto Encoder& world
cup 98 (WC98)  and Clark net traces & Python & Power, No. of requests   & - &  79\% error reduction  over canonical prediction \\ \hline 
		\end{tabular}}

\end{table*}

\section{Hybrid Prediction Models}
\begin{figure}[!htbp]
	\centering
	\includegraphics[width=0.9\linewidth]{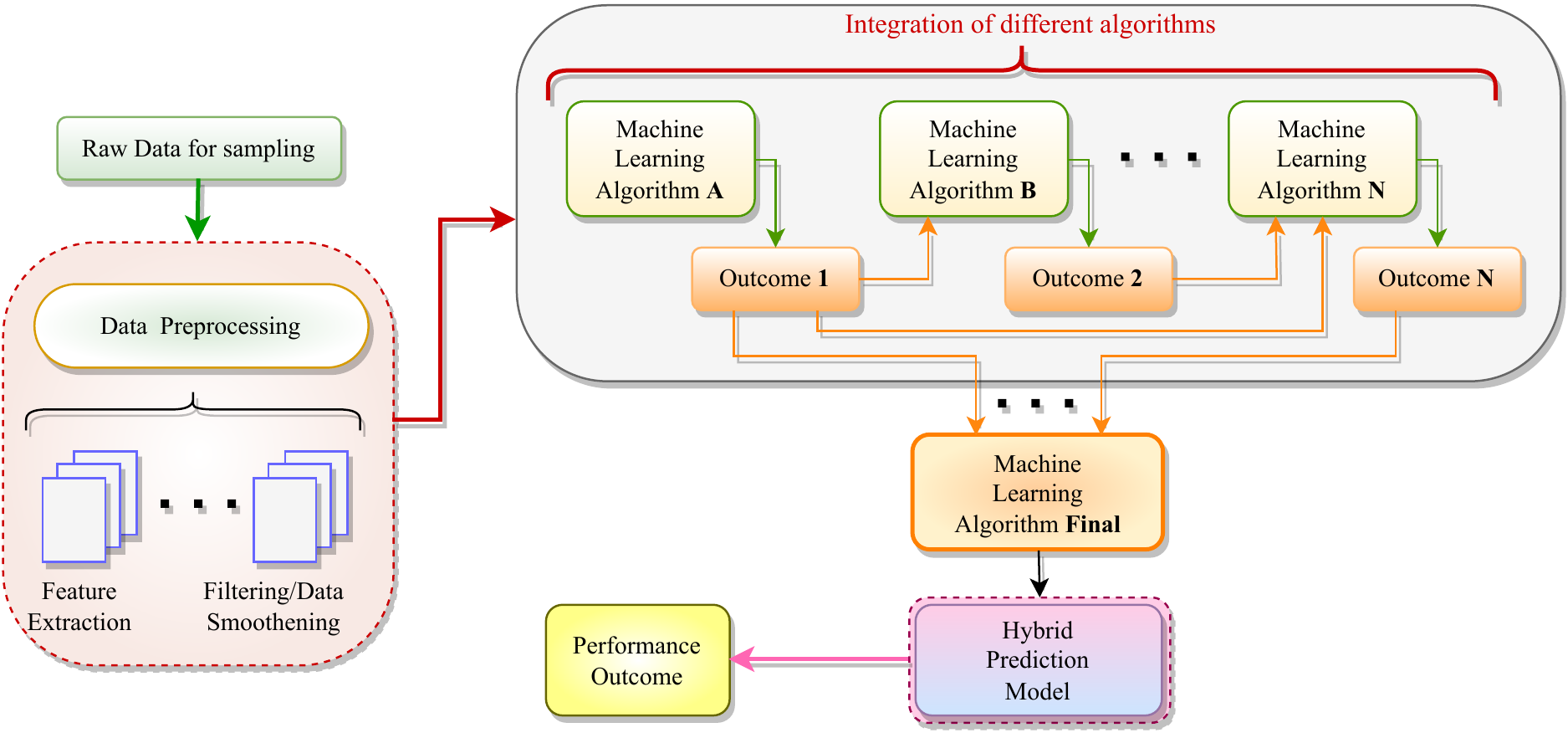}
	\caption{Hybrid Prediction Model}
	\label{fig:hybrid}
\end{figure}

The prediction models based on hybrid machine learning combines different machine learning algorithms (say $N$ algorithms) and  feed the outcome of one algorithm to another (one-way) to create an efficient  machine learning model for precise and  accurate predictions. These hybrid models are build using various collaborations such as `\textit{Classification} + \textit{Classification}'; `\textit{Classification} + \textit{Clustering}'; `\textit{Clustering} + \textit{Clustering}'; `\textit{Clustering} + \textit{Classification}'; `\textit{Classification} + \textit{Regression}'; `\textit{Clustering} + \textit{Regression}' etc. These combinations are selected as per the requirement and challenges respective to the intended problem to minimize features noise and biasness, reduce variance, and enhance the accuracy of prediction. A schematic representation of hybrid prediction model is depicted in Fig. \ref{fig:hybrid}. 
In this model, the raw and complex data samples are pre-processed by filtering and smoothening the data samples via extraction of significant features,   aggregation,  and scaling or normalization. The pre-processed data samples passes through a series of $N$ machine learning algorithms feeding their outomes as an input to other machine learning algorithm, generating a hybrid prediction model for performance measurement and deployment for the intended purposes. The significant key contributions related to hybrid approach based workload prediction models are discussed below.   \par
Kardani et al. \cite{kardani2020adrl} have developed  a hybrid Anomaly-aware Deep Reinforcement Learning-based Resource Scaling (ADRL) for dynamic resource scaling in the cloud environment. It presents an anomaly detection method for Deep Reinforcement Q-Learning-based decision making scheme that identifies anomalous states in the system and triggers actions accordingly. This work includes two levels of global and local decision-makers  to govern the necessary scaling actions. ADRL improved the QoS with essential actions only and increased stability of the system. 
A  hybrid Recurrent Neural Network (RNN) based
prediction model named `BHyPreC' is proposed in \cite{karim2021bhyprec} which comprises of Bidirectional Long Short-Term Memory (Bi-LSTM) on top of the stacked LSTM and GRU for prediction of VM resource usage. 
It improves the non-linear data analysis capability of Bi-LSTM, LSTM, and GRU models separately and confirmed better accuracy compared with other statistical models. It uses combined grid search technique of historical window size to optimize the model and determine the best possible set of window size. Bi et al. \cite{bi2021integrated} proposed integrated deep learning method named `BG-LSTM' which incorporates BiLSTM and GridLSTM  to achieve high-quality prediction of workload and resource time series. During preprocessing, it applies a  filter of Savitzky-Golay (SG) to reduce the standard deviation before smoothing workload. It can effectively extract complex and non-linear features of relatively longer time series and achieve high prediction accuracy. A Hierarchical Pythagorean Fuzzy Deep Neural Network (HPFDNN) is proposed in \cite{chen2019prediction} to predict the amount of cloud resources requirement. The neural representations of original sampling data are used as a supplementary approach for clear interpretations of true results which is  beyond the use of fuzzy logic. The users can determine the expected quantity of cloud services utilizing the forecasts of the deep neural network which will help in reducing cost. 
\par A hybrid autonomous  resource provisioning model based on MAPE-k control loop  for multi-tier applications is presented in \cite{khorsand2018fahp}. It is a combination of  the Fuzzy Analytical Hierarchy Process approach named as ` FAHP'. The
experimental results indicate that the proposed solution outperforms in terms
of allocated virtual machines, response time, and cost compared with the other
approaches. An Adaptive Classified Prediction Scheme (ACPS) is proposed in \cite{liu2017adaptive} which first categorises the workloads into different classes that are automatically assigned for different prediction models according to workload features. Further, the problem of the workload classification is transformed into a task assignment by establishing a mixed 0–1 integer programming model which is solved quickly by utilizing an improved branch and bound algorithm. Peng et al. \cite{peng2018multi} have   applied a GRU based Encoder-Decoder network named `GRUED'  containing two Gated
Recurrent Neural Networks (GRNNs) to address these issues. It has been evaluated via experiments  for the prediction of multi-step-ahead host workload in cloud computing. Shuvo et al.  \cite{shuvo2020lsru} have proposed a novel hybrid-method named `LSRU' for improving the prediction accuracy. LSRU is an integration of  LSTM and GRU  for short-time ahead prediction along with long-time ahead prediction with sudden burst of
workload.

\begin{table*}[!htbp]
	\caption{{Comparative Summary of Hybrid Machine Learning-based Prediction Models}}
	\label{summaryHybrid}
	\centering
	\resizebox{0.99\textwidth}{!}{
		\begin{tabular}{p{2.0cm} p{1.5cm}  p{6.5cm} p{2.5cm}p{2.04cm}  p{1.8cm} p{2.0cm} p{3.0cm}} \hline
			{\textbf{Notable Contributors (Timeline)} }&	\textbf{Model/ Approach/ Framework}&{\textbf{Workflow/ Strategy}}&\textbf{Datasets}&\textbf{Implementation/ Simulation tool}&{\textbf{Predicted parameters}}&{\textbf{Error metrics}}&{\textbf{Results or Remarks}} \\ \hline \hline
			Kardani et al. \cite{kardani2020adrl} (2021)&  ADRL: Deep RL+Q-Learning&  Deep Q-learning based RL model to respond CPU and memory bottleneck problem that help in resource scaling and decision making   & Web-based Rice University Bidding System (RUBiS)& Python with Java-based CloudSim&CPU, Memory, Response-time& MSE & improved QoS and stability \\ \hline

			Karim et al. \cite{karim2021bhyprec} (2021) & BiHyPrec: Bi-LSTM + LSTM + GRU & Pre-processed data is given to collaborative model of Bi-LSTM, LSTM,
			and GRU units that accomplishes a deep learning-based approach
			to  effectively tackle the complexity and non-linearity of time series data  & Bitbrains traces & Python and Google Colaboratory & CPU usage & MSE, MAPE, MAE, RMSE & performs better over ARIMA, LSTM, GRU, Bi-LSTM \\ \hline 
			
				Bi et al.  \cite{bi2021integrated} (2021) & BG-LSTM: Bi-LSTM + GridLSTM & Savitzky-Golay (SG)is used to smoothen data which is passed to hybrid model of Bi-LSTM and GridLSTM  for prediction &   Python & Google Cluster traces &CPU  and RAM usage & MSE, RMSLE, $R^2$ &  better accuracy over  SG-LSTM, SG-Bi-LSTM, SG-GridLSTM  \\ \hline 
				
				Chen et al. \cite{chen2019prediction} (2021) & HPFDNN: Hierarchical Pythagoras + Fuzzy  DNN &  Pythagorean fuzzy logic, neural
				representations, and deep neural network are integrated and network training method with adaptive learning rate is adopted to minimize the cost of cloud services for users 
			  &  Carnegie Mellon University dataset & Not mentioned & Number of requests & Accuracy cost and total cost & saved 202.48 dollars for real-time requests over existing methods \\ \hline
			  
			  Khorsand et al. \cite{khorsand2018fahp} (2018) &  FAHP &   FAHP and SVR algorithms are developed for workload prediction and resource provisioning and  the appropriate autoscaling  decisions   &  Clark Net, NASA, Synthetic workload & CloudSim and Open source RUBIS & Request arrival, response time, cost &  MSE, NMSE, RMSE, RMSSE & Reduces cost of rental resources for cloud service provider with QoS \\ \hline
			  Liu et al. \cite{liu2017adaptive} (2017) & ACPS  & Distinct  prediction models are automati-
			  cally assigned depending on the workload features & Google  Cluster traces & Python & Response time and Cost & Mean error, mean relative prediction error (MRPE) & Prediction error is reduced by 40.86\% over Linear Regression \\ \hline
			  
			 Peng et al.  \cite{peng2018multi} (2018) & GRUED & GRU encoder maps a
			 variable-length  workload sequence to a fixed-length
			 vector, and the GRU decoder maps the vector representation back to a
			 variable-length future workload series    & Google  Cluster and  Dinda workloads & Python, UNIX system  & CPU, Job arrivals & MAE, MAPE, RMSE, root mean segment squared error (RMSSE) & Reduce prediction error over LSTM \\ \hline
			 
			 Shuvo et al. \cite{shuvo2020lsru} (2020) & LSRU: LSTM + GRU & Some statistical methods such as AR, ES, ARMA, and
			 ARIMA for forecasting  and passed to LSTM and GRU combined unit for an improved prediction accuracy    & Bitbrains & Kaggle & CPU, Disk, memory, bandwidth &  MAE, MAPE, RMSE, MSE & Reduce prediction error over LSTM  and GRU  \\ \hline
		\end{tabular}
	}
\end{table*}

\section{Ensemble Prediction Models}
\begin{figure}[!htbp]
	\centering
	\includegraphics[width=0.8\linewidth]{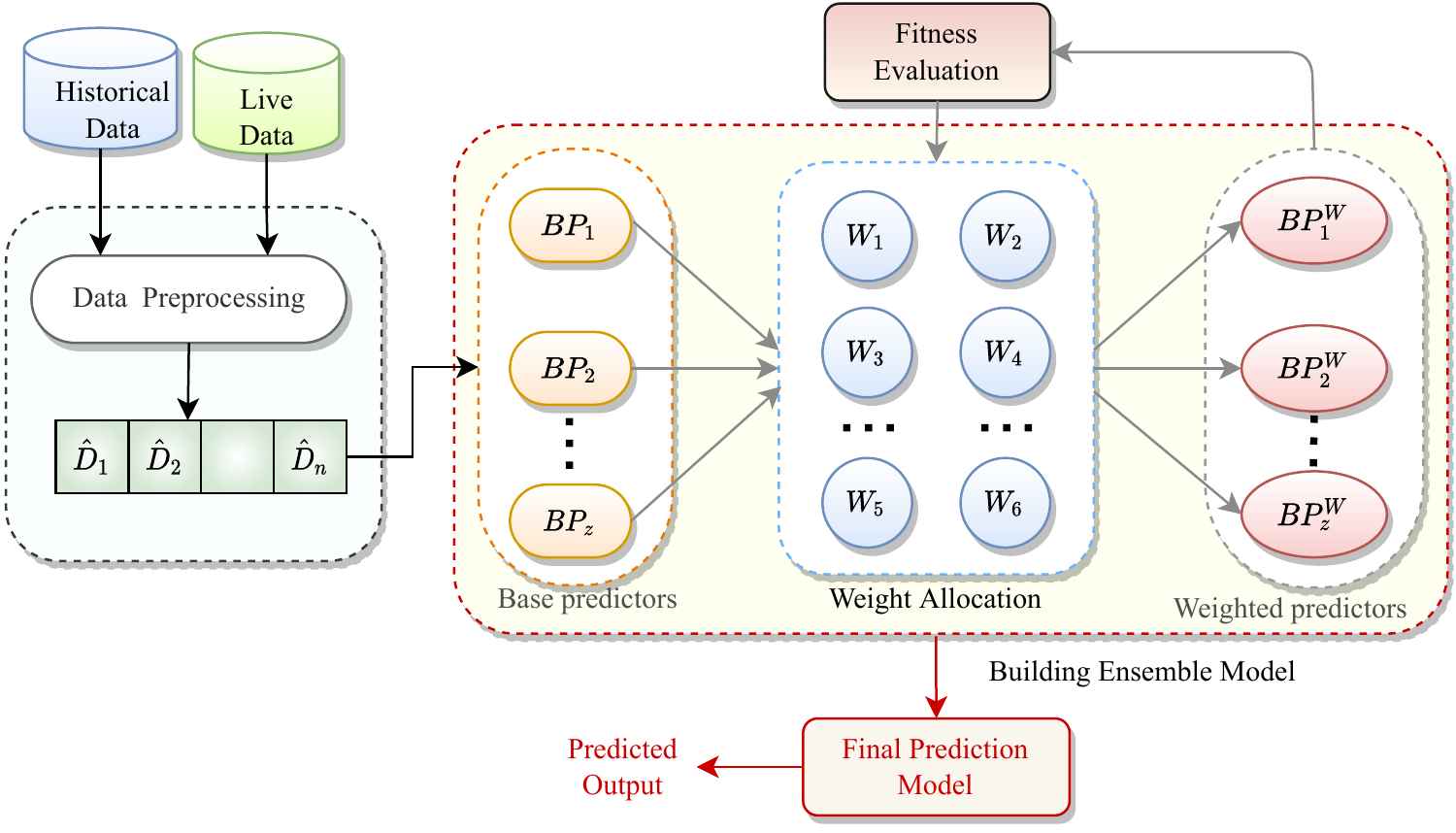}
	\caption{Ensemble Prediction Model}
	\label{fig:ensemble}
\end{figure}

An ensemble approach involves the use of multiple `\textit{Base Prediction}' ($BP$)
models or `\textit{Experts}' to forecast the expected future outcome of an event. The final outcome of an ensemble model is computed by combining
the forecasts of each expert using a voting engine. The conceptual
architecture of an ensemble based predictive approach is illustrated
in Fig. \ref{fig:ensemble}. 
The historical and live data samples are pre-processed and a `sliding window' is prepared and the input data vector thus created is given as input to all the base predictors \{$BP_1$, $BP_2$, ..., $BP_z$\}, where $z$ is the number of base predictors used for ensemble learning. The estimated outcome of each base predictor is assigned a weight value indicating their significance in the final prediction outcome. The weight allocation and their updation requires a learning process using suitable  optimization  method such as multi-class regression, priority based method, or evolutionary learning algorithm. The ensemble learning approach is more effective over the individual prediction method that suffers from challenges like, high variance, low accuracy, feature noise and bias. Basically, in ensemble approach, the various machine learning models work independently of each other to give a prediction and  a voting system (hard or soft voting) based on weighted values associated to each base predictor determines the final prediction. The prominent models based on ensemble learning  are discussed below.
 \par 
Singh et al. \cite{singh2014ensemble} have addressed the problem of  
extensive range of workloads prediction by extending and adapting two online
ensemble learning methods including  Weighted Majority (WM) and Simulatable Experts (SE). The classical SEs are extended from binary outcome space to k-outcome space to make them suitable for solving any k-class problem (designated as `kSE'). The Weighted Majority ensemble model parameters are regenerated incrementally  making these algorithms computationally more efficient and suitable for  handling massive range of online data streams (designated as `WMC'). These models are evaluated using large datasets of 1570 servers and have verified that approximately 91\% servers can be correctly predicted with the extended versions of these algorithms.   
Feng et al. \cite{feng2022fast} have proposed an ensemble model for Forecasting workloads with Adaptive Sliding window and Time locality integration named FAST.  An adaptive sliding window algorithm is developed considering  correlations of trend and time, and random fluctuations of forthcoming workloads  to maximize accuracy of prediction with lower overhead. Also,  a time locality concept for local-predictor
behavior is accomplished for the error-based integration strategy. The entire model is integrated by developing  a multi-class regression weighting algorithm. The performance of the model is validated using Google Cluster trace datasets.
 An integrated model for temporal prediction of workloads is proposed in \cite{bi2019temporal} which combines  Savitzky–Golay (SG) filter and wavelet
 decomposition with stochastic configuration networks to predict
approaching workload. In this model, a task time series is smoothened using SG filter and decomposed into components via  wavelet
decomposition method. This model named  `SGW-S' is able to  characterize the statistical
features of both trend and detailed components and achieved an improved performance with faster learning. 

A cloud resource forecasting model named `CloudInsight' (ClIn) based on ensemble prediction approach is proposed in \cite{kim2020forecasting}. This model meticulously locates the most admissible machine learning prediction approach by training a statistical features based classifier for the accurate estimation of job arrivals. It employs a number of local predictors or experts and builds an ensemble prediction model using them by dynamically determining the significant weights (or contributions) of each local predictor. The adaptive
weight scores are optimized at regular intervals with the help
of multi-class regression with a SVM classifier for selection of
the most appropriate prediction model with highest accuracy
during respective prediction interval. This model is tested using three different categories of workloads including: web, cluster and high performance computing workloads to prove its efficacy over existing prediction models. 
Baig et al. \cite{iqbal2019adaptive} have proposed an ensemble and adaptive  cloud resource estimation model named 'Adaptive Model Selector' (AMS). This model 
determines the most admissible machine learning prediction
approach by training a statistical features based classifier
for the accurate estimation of resources utilization. It builds a classifier using Random Decision Forest (RDF) to predict the
best model for a given sliding window data by training and re-training periodically at regular time-intervals. The selected features and identified prediction methods
are logged as training data and passed to sliding window for learning process. The performance of this model is evaluated using Google Cluster, Alibaba, and Bitbrains datasets. Kumar et al. \cite{kumar2020ensemble} have presented an ensemble learning based workload forecasting  model named `E-ELM'  that uses extreme learning machines and their associated forecasts are weighted by a voting engine. In this work, a metaheuristic algorithm inspired by blackhole theory  is applied to decide the optimal weights. The
accuracy of the model is evaluated for CPU and memory demand requests of Google cluster traces and for CPU utilization of PlanetLab VM traces. 
Chen et al. \cite{chen2015self} have proposed a self-adaptive prediction method using ensemble model and subtractive-fuzzy clustering based fuzzy neural network (ESFCFNN). The user preferences and demands are characterized and an ensemble prediction model is constructed using several base predictors.

\begin{table*}[!htbp]
	\caption{{Comparative Summary of Ensemble Learning-based Prediction Approaches}}
	\label{summaryEnsemble}
	\centering
	\resizebox{0.99\textwidth}{!}{
	\begin{tabular}{p{2.0cm} p{1.5cm}  p{6.5cm} p{2.5cm}p{2.0cm}  p{1.8cm} p{2.0cm} p{3.0cm}} \hline
		{\textbf{Notable Contributors (Timeline)} }&	\textbf{Model/ Approach/ Framework}&{\textbf{Workflow/ Strategy}}&\textbf{Datasets}&\textbf{Implementation/ Simulation tool}&{\textbf{Predicted parameters}}&{\textbf{Error metrics}}&{\textbf{Results or Remarks}} \\ \hline \hline
			Singh et al. \cite{singh2014ensemble} (2014) & kSE+WMC  &SE is extended to k-outcome space and WM is improved for incremental and computationally efficient learning process& Dataset G(1570)&Not mentioned  & server workload & MAPE &
		  89\% accurate predictions as compared with 13–24\% for baseline algorithms  \\ \hline
			
	Feng et al.	\cite{feng2022fast} (2022) &	FAST &  The adaptive sliding window considers all types of workload trends with time locality concept for error-based integration is developed to enhance prediction accuracy&Google  Cluster traces   & Python & CPU, memory & Absolute Error, RMSE, $R^2$ & improved prediction accuracy by 14.99\% to 27.55%
	on RMSE and 22.57\% to 76.86\% on $R^2$ \\ \hline
	Bi et al.	\cite{bi2019temporal} (2019)&	SGW-S  & SG filter and wavelet
			decomposition is integrated with stochastic configuration networks to predict
			 workload  with high accuracy &Google Cluster traces    & Not mentioned   &task arrival rate & MSE, $R^2$& Use of SG-filter with wavelet decomposition helps improve the prediction accuracy  \\ \hline
Kim et al.	\cite{kim2020forecasting} (2020)&	ClIn: ensemble with multi-class regression &Different local predictors constitutes an ensemble prediction model by dynamically determining and updating the significant weights  of each  predictor & Web, cluster, HPC Grid workloads & Python  & Job arrival per unit time & Normalised RMSE, Absolute Error & Up to 15\%–20\% less
under-/over-provisioning with high cost-efficiency and low SLA violations\\ \hline
		
Baig et al. \cite{iqbal2019adaptive} (2019) &		AMS  & Predict the forthcoming resource demand with the best predictor model as selected by the RDF classifier which is updated periodically with time&Alibaba, Bitbrains, Google Cluster  &  Python     & CPU & RMSE, MAE  &  improved prediction accuracy from 6\%-27\% over current methodologies\\ \hline 
Kumar et al. \cite{kumar2020ensemble} (2020) &	E-ELM  & ELM based local predictors are trained and selected by employing a weight updation method with a help of a metaheuristic algorithm  &MATLAB   & Google Cluster and PlanetLab traces  & CPU, memory & RMSE, MAE & RMSE reduced up to 99.20\% over existing methods \\ \hline	
		
		Chen et al. \cite{chen2015self} (2015) & Subtractive-fuzzy clustering&   user preferences and demands are characterized into an ensemble prediction model subtractive-fuzzy clustering based fuzzy neural network & Data Flow Statistics traces& CloudSim & Network  resource traces& MSE, MAE & Effective in resource demand prediction \\ \hline
		\end{tabular}
	}
\end{table*}

\section{Quantum Neural Network based Prediction Models}

\begin{figure}[!htbp]
	\centering
	\includegraphics[width=0.7\linewidth]{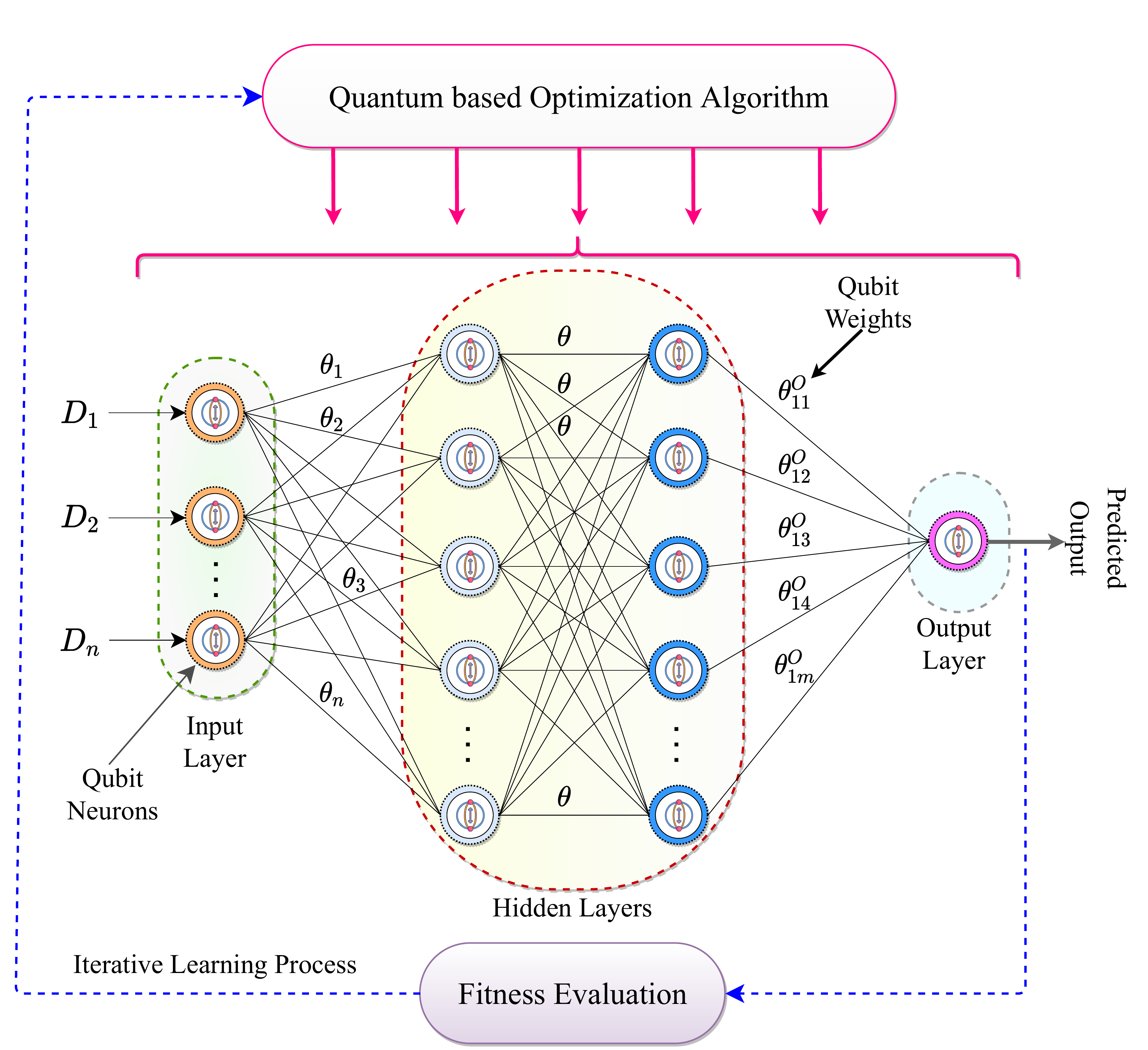}
	\caption{Quantum Neural Network}
	\label{fig:qnn}
\end{figure}

A Quantum Neural Network (QNN) based prediction model is an intelligent  model for prediction with the machine learning capabilities of neural network and computational proficiency of quantum mechanics to achieve  prediction accuracy with high precision.  Basically, it is a neural network comprising of Qubit neurons and Qubit weights instead of real-numbered values and the training data information is also propagated in the form of Qubits.   A Qubit is represented as a one, a zero, or any quantum superposition of these states. Mathematically, it can be realized as: $|\Psi\textgreater = \alpha|0\textgreater +  \beta|1\textgreater  $, where $\alpha$ and $\beta$ are complex numbers specifying the probability amplitudes of states $|0\textgreater$ and $|1\textgreater$ respectively. The schematic representation of QNN based prediction model is shown in Fig \ref{fig:qnn} which comprises of one input, multiple hidden and one output layers having $n$, $p$ and $q$ qubit nodes, respectively which represents $n$-$p$-$q$ Qubit network architecture.  The inter-connection weights between Qubit neurons of different layers are also taken in the form of Qubits that are adjusted during learning process.
The state transitions of Qubit neurons are derived from the various Quantum gates such as Rotation gate, Controlled-Not gate etc. The training data samples values are extracted and aggregated into a specific time-interval such as 5 minutes which are scaled in a specific range using a normalization function. The normalized data values are transformed into  quantum state values or Qubits  by applying the effect of qubit rotation using  Eqs. (\ref{eq6}) and (\ref{eqq});
\begin{gather}
{y^{In}}_i=f({\Theta_i}^{In})\label{eq6}\\              
\Theta_i=\frac{\pi}{2}\times \mathcal{D}_i \label{eqq} 
\end{gather}  
where $\mathcal{D}_i$ is the $i^{th}$ input data point, $\Theta_i$ is $i^{th}$ Quantum input point to the network. The QNN model extracts relevant and significant patterns from actual workload  and analyzes $n$ previous workload values to estimate forthcoming workload information at the next $(n+1)^{th}$ instance of time within the datacentre. The learning process of  QNN is derived by a Qubit based optimization algorithm that can manipulate, explore, and exploit  Qubits to regenerate the population and prevent the problem of stagnation. The workload prediction methods using QNN is still at an infancy stage.  
\par
The only work established thus far using QNN based model for workload prediction is proposed in \cite{singh2021quantum}. This model  exploits the computational efficiency of quantum computing by encoding workload information into Qubits and propagating this information via network for estimatimation of the workload or resource demands with enhanced accuracy proactively. The rotation and reverse rotation effects of the C-NOT gate served the activation function at the hidden and output layers to optimize the Qubit weights.  Self Balanced Adaptive Differential Evolution (SB-ADE) algorithm is developed to optimize qubit network weights. This model is evaluated using three different categories of workloads where the prediction accuracy is substantially is improved over the existing approaches.  A  workload prediction model using complex numbers,  is presented in  \cite{qazi2018cloud} where a  high capability of learning and better accuracy is applied to  multi-layered neural networks with multi-valued neurons (MLMVN) prediction model in less time.
\section{Performance Evaluation }

\subsection{Experimental Set-up}
The simulation experiments are conducted on a server machine assembled with two Intel\textsuperscript{\textregistered} Xeon\textsuperscript{\textregistered} Silver 4114 CPU with a 40 core processor and 2.20 GHz clock speed. The computation machine is deployed with 64-bit Ubuntu 16.04 LTS, having 128 GB RAM. {All the selected best prediction works based on Evolutionary Neural Network (ENN), Ensemble Learning (EL), Hybrid Learning (HL), Deep Learning (DL), Evolutionary Quantum Learning   were implemented  in Python 3.7 with the details of various intended parameters with their values listed in Table \ref{table:name1}.}   

\begin{table}[!htbp]
	\centering
	\caption{Experimental set-up parameters and their values}
	\begin{tabular}{|p{1.1cm}|l|p{3cm}|} \hline
\textbf{Preiction Models}	& \textbf{Parameter} & \textbf{Values}     \\ \hline
		\multirow{8}{*}{\rotatebox[origin=c]{90}{\shortstack{Evolutionary \\ Neural Network}}}
		&  Input nodes   & 10 \\
		&  Hidden nodes             & 7 \\
		& Maximum iteration         & 250 \\
		& Training data size        & 70\%  \\
		& Mutation learning period  & 10 \\
		& Crossover learning period & 10 \\
		& Size of population & 15 \\
		& Training algorithm       & Differential Evolution \\ \hline
		\multirow{8}{*}{\rotatebox[origin=c]{90}{\shortstack{Ensemble \\ Learning }}}
		 & \#ELM networks & [10,100] \\
		&  Input nodes              & [7,70] \\
		&  Hidden nodes             & [5,50] \\
		& Population size           & 20 \\
		& Maximum iteration         & 100  \\
		& Accuracy threshold        & 0.007 \\
		& Training data size        & 70\% \\
		& Training algorithm       & Blackhole  Optimization \\ \hline            
		\multirow{8}{*}{\rotatebox[origin=c]{90}{\shortstack{Hybrid \\ Learning }}} 
		&  Input nodes  & 100 \\
		&  Epochs             & 20-50 \\
		& Batch size           & 16 \\
		& Activation function         & tan$h$  \\
		& Training data        & 70\% \\
		& Number of epochs    & 500-1000 \\
		& Batch size & 1-4 \\
		& Training algorithm       & Adam Optimizer\\ \hline
		\multirow{8}{*}{\rotatebox[origin=c]{90}{\shortstack{Deep \\ Learning}}}
		 &  Deep learning libraries:& \\
		& tensorflow           & 0.12.1 \\
		& keras         & 1.2  \\
		& Training data        & 70\% \\
		& Number of epochs    & 500-1000 \\
		& Neurons      & 4-10 \\
		& Batch size & 1-4 \\ 
		& Training algorithm       & Gradient descent \\\hline            
		\multirow{9}{*}{\rotatebox[origin=c]{90}{\shortstack{Quantum \\ Learning}}}
		&  Input nodes & 10 \\
		& Hidden nodes             & 7 \\
		& Output nodes         & 1 \\
		&Number of epochs         & 50  \\
		& Training data        & 70\% \\
		& Population size           & 15 \\
		& Mutation learning period  & 5 \\
		& Crossover learning period & 5 \\
		& Training algorithm       & Quantum Differential Evolution \\ \hline      
	\end{tabular}
	
	\label{table:name1}
\end{table}

\subsection{ Data Sets }
The performance analysis and comparison of various machine learning based prediction models are executed using three different benchmark datasets including CPU and memory usage traces from  Google Cluster Data (GCD) \cite{Reiss2011} and CPU usage from PlanetLab (PL) virtual machine traces \cite{beloglazov2012optimal}. GCD workload provides behavior of cloud applications for the cluster and big data analytics such as Hadoop which gives resource: CPU, memory, and disk I/O  request and usage information of 672,300 jobs executed on 12,500 servers  collected over a period of 29 days. The CPU and memory utilization  percentage of VMs are obtained from the given CPU and memory usage percentage for each job in every five minutes over a period of twenty-four hours. PL contains CPU utilization of more than 11000 VMs measured every five minutes during ten random days in March-April, 2011. The respective values of resource usage are extracted and aggregated according to different prediction-intervals such as 5, 10, 20, ...., 60 minutes. These values are re-scaled in the range [0, 1] using the normalization formula stated in Eq. (\ref{normalisation}). Table \ref{table:(NASA)workloadcharacteristic} shows the statistical characteristics of the evaluated workloads. 

\begin{table}[!htbp] 	
	\caption[Table caption text] {Characteristics of evaluated workloads  }  
	\label{table:(NASA)workloadcharacteristic} 	
	\centering
	\resizebox{9cm}{!}{
		\begin{tabular}{    l  l  l   l  l  }
			\hline
				\textbf{ Workload} &\textbf{ Duration} & \textbf{Jobs}& \textbf{Mean(\%)} & \textbf{St.dev} \\
			
			\hline
			GCD-CPU ($G^C$) & 10 days&2 M &21.84&13.62\\
			GCD-memory ($G^M$) & 10 days& 2 M& 19.55&16.6 \\
			PL-CPU ($P^C$) & 10 days	& 1.5 M& 19.77&14.55\\	
			
			\hline
	\end{tabular}}
\end{table}
\subsection{Evaluation Metrics}
Forecast accuracy of the prediction models  are evaluated using following error metrics:

\textit{Mean Squared Error ($\mathcal{MSE}$)}:
It is one of the well known metric to measure the accuracy of prediction models, which puts high penalty on large error terms. The model is considered to be more accurate if its score is closer to zero. The mathematical representation of the metric is mentioned in Eq.(\ref{eq:mse}), where $m$ is the number of data points in the workload trace, $\mathcal{Z}^{Ac}(t)$ and $\mathcal{Z}^{Pr}(t)$ are actual and predicted workload values, respectively at $t^{th}$ instance. 

\begin{equation}\label{eq:mse}
\mathcal{MSE} = \frac{1}{m} \sum_{t=1}^{m} (\mathcal{Z}^{Ac}(t) - \mathcal{Z}^{Pr}(t))^2
\end{equation}

\textit{Mean Absolute Error ($\mathcal{MAE}$)}:
In mean squared error the square of higher error values may receive more weightage which can effect the accuracy of prediction. While $\mathcal{MAE}$ assigns equal weight to each error component and measures the accuracy of the prediction model by computing the mean of absolute differences between
actual ($\mathcal{Z}^{Ac}(t)$) and predicted ($\mathcal{Z}^{Pr}(t)$) workloads at $t^{th}$ time-instance as shown in Eq.~\eqref{eq:mae}. It produces a non negative number to evaluate the forecast accuracy and if it is close to zero, forecasts are very much similar to actual values.

\begin{equation}\label{eq:mae}
\mathcal{MAE} = \frac{1}{m} \sum_{t=1}^{m} |\mathcal{Z}^{Ac}(t) - \mathcal{Z}^{Pr}(t)|
\end{equation}
\subsection{Results}
The performance of different prediction models including Evolutionary Quantum Neural Network (EQNN) \cite{singh2021quantum}, Ensemble Learning (EL) \cite{kumar2020ensemble}, Hybrid Learning \cite{shuvo2020lsru}, Deep Learning (DL) \cite{kumar2018long}, Evolutionary Neural Network (ENN) \cite{kumar2018workload}; are thoroughly investigated and compared using extensive range of heterogeneous cloud applications and variable resource utilization by VMs. We have evaluated and compared the different types of learning-based models for $\mathcal{MSE}$ with confidence metrics, $\mathcal{MAE}$, Absolute Error Frequency (AEF), and time elapsed in Training (TT). 

\subsubsection{Mean Squared Error}
\begin{figure*}[!htbp]
	
	\centering
	
	\subfigure[GCD-CPU traces]{\includegraphics[width=.315\textwidth]{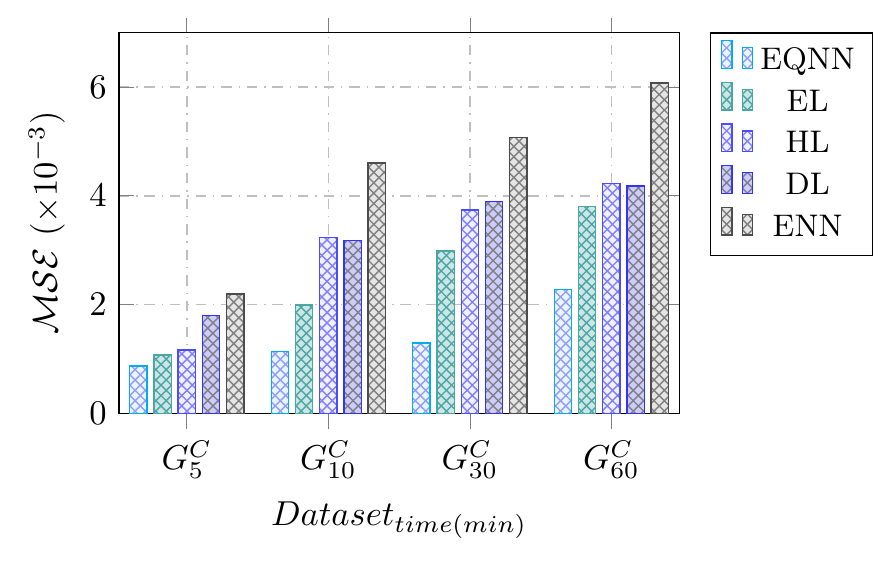}}\hfill
	\subfigure[GCD-Memory traces  ]{\includegraphics[width=.315\textwidth]{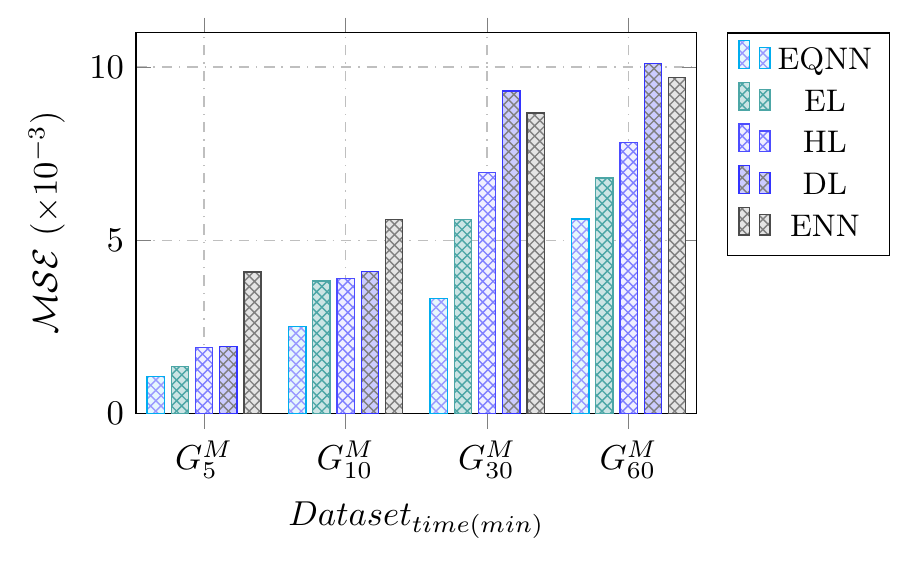}}\hfill
	\subfigure[PlanetLab CPU traces ]{\includegraphics[width=.315\textwidth]{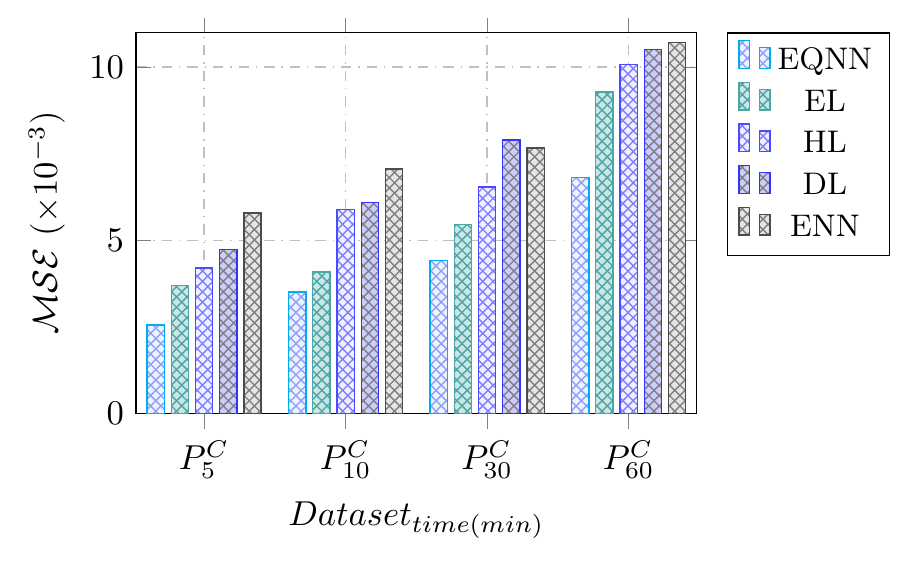}}	
	\caption{Mean Squared Error }
	\label{fig:mse}
	
\end{figure*}

 Fig. \ref{fig:mse} compares mean  values of three different categories of workloads including GCD-CPU traces (Fig. \ref{fig:mse}a), GCD-Memory traces (Fig. \ref{fig:mse}b), and PL-CPU traces (Fig. \ref{fig:mse}c). It is observed from the figures that $\mathcal{MSE}$ varies differently with the variety of prediction models and increases with the size of prediction interval from 5 to 60 minutes. The is due to the fact that with increment in the size of prediction window, the number of avaiable training data samples decreases. The prediction accuracy with respect to reduction in the values of average $\mathcal{MSE}$ follows a trend: EQNN $<$ EL $\leq$ HL $<$ DL $<$ ENN. Also, it is notified that the difference among prediction errors ($\mathcal{MSE}$) is lesser for shorter prediction interval which increases with the decrement in the number of training data samples with growing prediction window-size. Hence, it can be concluded that for short-term prediction, all types of prediction models produce an expected level of prediction accuracy, and the major difference in the performance of prediction arises with the long-term prediction intervals.  The obtained experimental values demonstrates that Quantum learning-based prediction model (EQNN) is providing least prediction error for majority of the prediction intervals and most of the data traces while evolutionary learning-based prediction model (i.e., ENN) produces highest $\mathcal{MSE}$ values for majority of the experimental cases.  For GCD-CPU (Fig. \ref{fig:mse}a), EQNN is performing consistently best among all the approaches because of the employment of Qubits and Quantum superposition effects which imparts precise and intuitive learning of correlations and relevant patterns during learning process. Moreover, EL  gives lesser $\mathcal{MSE}$ values than HL, DL, and ENN because of involvement of multiple base prediction models which adaptively selects the best prediction model each time and rejuvenates the learning process by updating  the weights associated with the respective respective base predictors. The prediction errors for HL is lesser as compared with DL by reason of the integration of filtering and smoothening approaches (by leveraging various filtering methods like SG-Filters or using GRU to improve the accuracy while minimizing the drawback of LSTM)  before actual prediction. In HL, two or more approaches combines cooperatively by diminishing the limitations of each intended approach and thus producing an effective prediction model to forecast resource usage of extensive range of cloud workloads. The experiments of DL are performed using LSTM based prediction method which performs better than ENN for all the cases including $G^{C}_{5}$ to $G^{C}_{60}$. The resultant graph shown in Fig. \ref{fig:mse}b reveals similar trends for the GCD-Memory traces where HL and EL show closer performance i.e., lesser than EQNN but superior than DL and ENN based prediction models. The results achieved using DL based prediction model entails improved accuracy in terms of lesser values of $\mathcal{MSE}$ than ENN for all the respective experiments  except for the cases of $G^{M}_{30}$ and $G^{M}_{60}$.  Fig. \ref{fig:mse}c represents the comparison of $\mathcal{MSE}$ for PL CPU traces, wherein the difference among prediction error values is slight but significant that supports the aforementioned trend of performance. Futhermore, the confidence metrics are computed for the achieved $\mathcal{MSE}$ results as shown in Table \ref{table:result}, wherein error margin (EM) and confidence-interval (CI) are reported for all the experimental cases.    

\begin{table*}[!htbp] 	
	\caption[Table caption text] {{Confidence metrics for Mean Squared Error} }  
	\label{table:result} 	
	\centering
	\resizebox{16cm}{!}{
		\begin{tabular}{  c  c  c  c   c c c c }
			\hline
			
			{\textbf{Dataset}}	& {\textbf{PWS}$^a$}& {\textbf{ Metrics}} & \textbf{ EQNN}& \textbf{EL} & \textbf{HL} & \textbf{DL} &\textbf{ENN} 
			\\
			
			\hline
			\multirow{8}{*}[-0.4ex]{\rotatebox{90}{\textbf{GCD-CPU}}}&\multirow{2}{*}{{5}} &EM &5.3012E-06 &1.3824E-06  &2.3666E-06 & 1.8460E-06 & 2.7320E-06\\ 
			&	&CI   & 8.799E-03 - 8.801E-03& 1.079E-03 - 1.080E-03  &  1.169E-03 - 1.170E-03& 1.799E-03 - 1.800E-03 &2.199E-03 - 2.200E-03 \\ \cline{2-8}
			
			&\multirow{2}{*}{{10}} &EM &1.7969E-06 & 3.6608E-06 &5.4890E-06 & 4.5940E-06 &3.2681E-06 \\ 
			&	&CI   & 1.398E-03 - 1.140E-03 & 1.992E-03 - 1.993E-03       & 3.239E-03 - 3.241E-03& 3.895E-03 - 3.900E-03  &4.599E-03 - 4.600E-03 \\ \cline{2-8}
			
			&\multirow{2}{*}{{30}} &EM & 2.1890E-06& 1.4865E-06 &6.8200E-05 & 4.9826E-06 & 6.8210E-06\\ 
			&	&CI   & 1.298E-03 - 1.299E-03&  2.989E-03 - 2.990E-03   & 3.733E-03 - 3.747E-03& 4.179E-03 - 4.180E-03   & 5.069E-03 - 5.071E-03  \\ \cline{2-8}
			
			&\multirow{2}{*}{{60}} &EM &1.1743E-06 &1.702E-06  &6.476E-06 & 5.8040E-06 & 8.2199E-06 \\ 
			&	&CI   & 2.280E-03 - 2.282E-03& 3.798E-03 - 3.801E-03&4.229E-03 - 4.231E-03& 4.179E-03 - 4,181E-03&2.39E-05 \\ \hline	\hline
			\multirow{8}{*}[-0.4ex]{\rotatebox{90}{\textbf{GCD-Memory}}}&\multirow{2}{*}{{5}} &EM &5.7890E-05 &6.7740E-06  &1.7841E-06 & 8.7341E-06 & 1.6839E-05\\ 
			&	&CI   & 1.079E-03 - 1.081E-03& 1.369E-03 - 1.370E-03  &  1.919E-03 - 1.920E-03& 1.9391E-03 - 1.941E-03 &4.090E-03 - 4.093E-03 \\ \cline{2-8}
			
			&\multirow{2}{*}{{10}} &EM &2.0799E-05 &4.0514E-06  &4.9616E-06 &6.6035E-06  & 2.5002E-05\\ 
			&	&CI   &2.507E-03 - 2.512E-03  &3.829E-03 - 3.830E-03 &3.909E-03 - 3.910E-03 &4.099E-03 - 4.101E-03  & 5.597E-03 - 5.603E-03\\ \cline{2-8}
			
			&\multirow{2}{*}{{30}} &EM &6.8200E-05 &6.4132E-05  & 7.1294E-05&5.813E-05  & 2.5785E-05\\ 
			&	&CI   &3.329E-03 - 3.331E-03 &5.593E-03 - 5.606E-03  & 6.953E-03 - 6.954E-03& 9.307E-03 - 9.310E-03 & 8.667E-03 - 8.673E-03\\ \cline{2-8}
			
			&\multirow{2}{*}{{60}} &EM & 4.8290E-05& 9.9990E-05  &1.5458E-05 &2.4786E-06  & 3.5760E-05 \\ 
			&	&CI   &5.615E-03 - 5.625E-03 &6.799E-03 - 6.801E-03 &7.818E-03 - 7.822E-03  & 10.0991E-03 - 10.0996E-03 & 9.696E-03 - 9.703E-03\\ \hline 			
			 \hline
			\multirow{8}{*}[-0.4ex]{\rotatebox{90}{\textbf{PlanetLab CPU}}}&\multirow{2}{*}{{5}} &EM &4.2810E-06 &4.6890E-06  & 4.6170E-06  & 6.8910E-06  & 2.3980E-06\\ 
			&	&CI   & 2.559E-03 - 2.560E-03& 3.699E-03 - 3.700E-03  & 4.199E-03 - 4.200E-03 &4.739E-03 - 4.740E-03  & 5.794E-03 - 5.795E-03  \\ \cline{2-8}
			
			&\multirow{2}{*}{{10}} &EM & 8.4210E-06& 9.8140E-06 & 7.1100E-06&  8.7642E-06 &1.0890E-05 \\ 
			&	&CI   &3.499E-03 - 3.510E-03 & 4.092E-03 - 4.093E-03 & 5.899E-03 - 5.900E-03 & 6.099E-03 - 6.101E-03 &7.058E-03 - 7.061E-03 \\ \cline{2-8}
			
			&\multirow{2}{*}{{30}} &EM & 6.2814E-05&8.6411E-05  &1.8230E-05 & 6.4210E-06 & 8.6810E-06\\ 
			&	&CI   &4.423E-03 - 4.436E-03 &5.451E-03 - 5.468E-03 &6.538E-03 - 6.541E-03 & 7.899E-03 - 7.900E-03 &7.669E-03 - 7.670E-03 \\ \cline{2-8}
			
			&\multirow{2}{*}{{60}} &EM &8.4210E-05 &9.9610E-05  & 6.4280E-05 & 2.8136E-05 & 9.8134E-05 \\ 
			&	&CI   & 6.811E-03 - 6.824E-03 & 9.270E-03 - 9.289E-03& 10.075E-03 - 10.088E-03&10.449E-03 - 10.450E-03  & 10.690E-03 - 10.709E-03\\ \hline 
					
	\end{tabular}}
	\\
	\footnotesize{\tiny{{$^a$ PWS: Prediction Window Size, EM: Error Margin, CI: Confidence Interval}}}	
\end{table*}

\subsubsection{Mean Absolute Error}
\begin{figure*}[!htbp]
	
	\centering
	
	\subfigure[GCD-CPU traces]{\includegraphics[width=.315\textwidth]{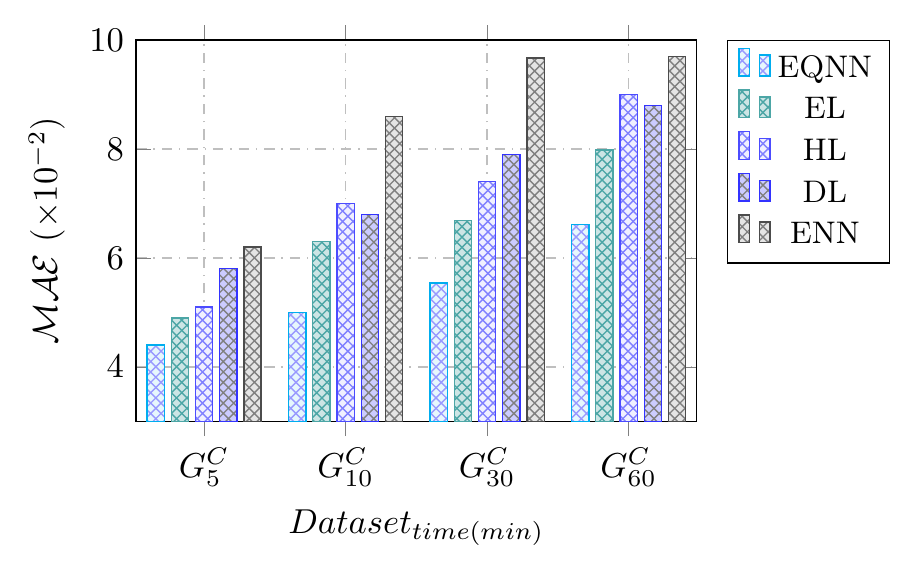}}\hfill
	\subfigure[GCD-Memory traces  ]{\includegraphics[width=.315\textwidth]{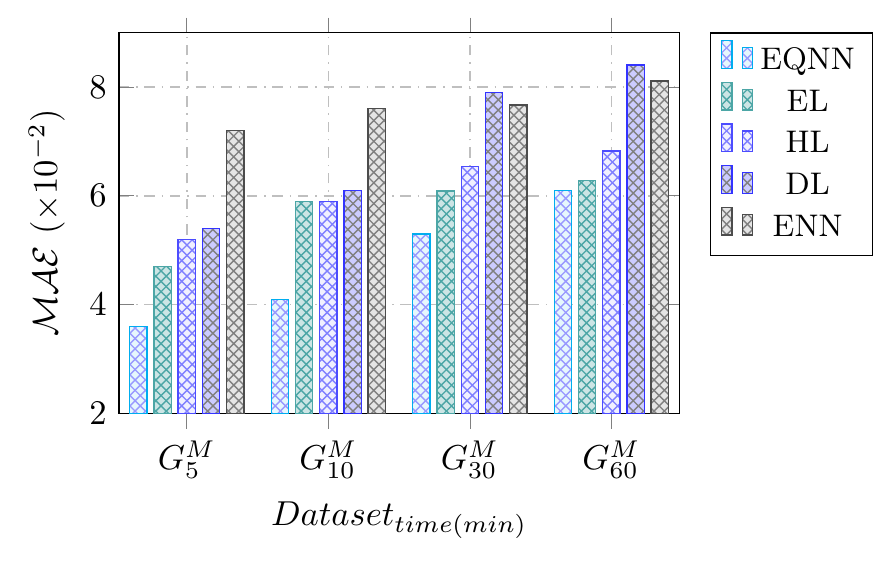}}\hfill
	\subfigure[PlanetLab CPU traces ]{\includegraphics[width=.315\textwidth]{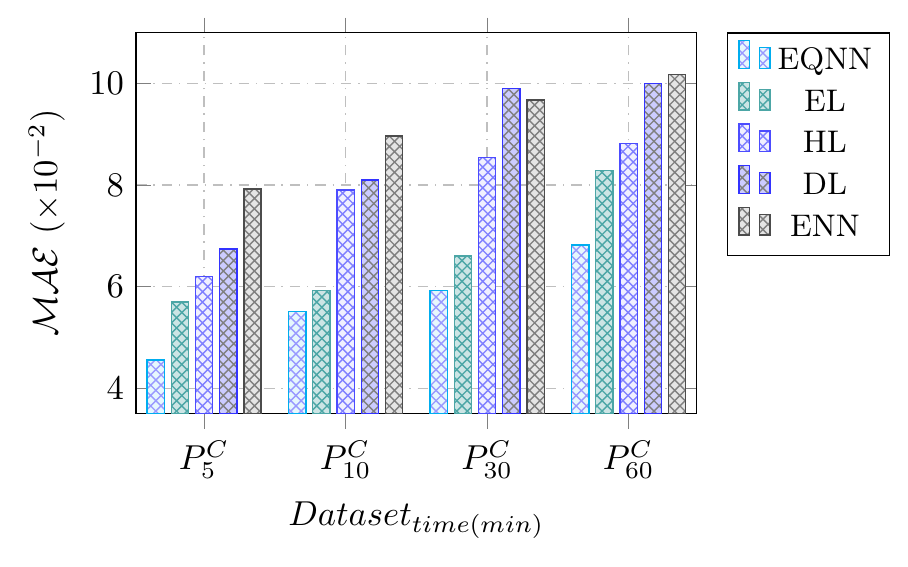}}
	
	\caption{Mean Absolute Error }
	\label{fig:mae}
	
\end{figure*}

The comparison of resultant values of $\mathcal{MAE}$ obtained for different prediction models over various datasets is  presented in Fig. \ref{fig:mae}. Similar to the $\mathcal{MSE}$, the $\mathcal{MAE}$ values follows the common trend of the performance for all the three evaluated workloads including GCD-CPU (Fig. \ref{fig:mae}a), GCD-Memory (Fig. \ref{fig:mae}b), and PL-CPU (Fig. \ref{fig:mae}c). As depicted in the three aforementioned consequent bar graphs, the  $\mathcal{MAE}$ values decreases in the order:  EQNN $<$ EL $\leq$ HL $<$ DL $<$ ENN. Further, it is observed that the difference among prediction errors ($\mathcal{MAE}$) is  increasing with the growing prediction window-size because of the decrement in the number of training data samples. Hence, it can be concluded that for short-term prediction, all the types of prediction models produces an expected level of prediction accuracy, and the major difference in the performance of prediction arises with the long-term prediction intervals. The reason behind this is that as the number of training samples decreases, there is not enough learning of the patterns and the respective prediction models begins underestimating the relevant information from the training dataset, resulting into a lesser ability of developing necessary correlations and performance degradation with diminished pattern learning.

\subsubsection{Training time}
\begin{figure*}[!htbp]
	
	\centering
	
	\subfigure[GCD-CPU traces]{\includegraphics[width=.315\textwidth]{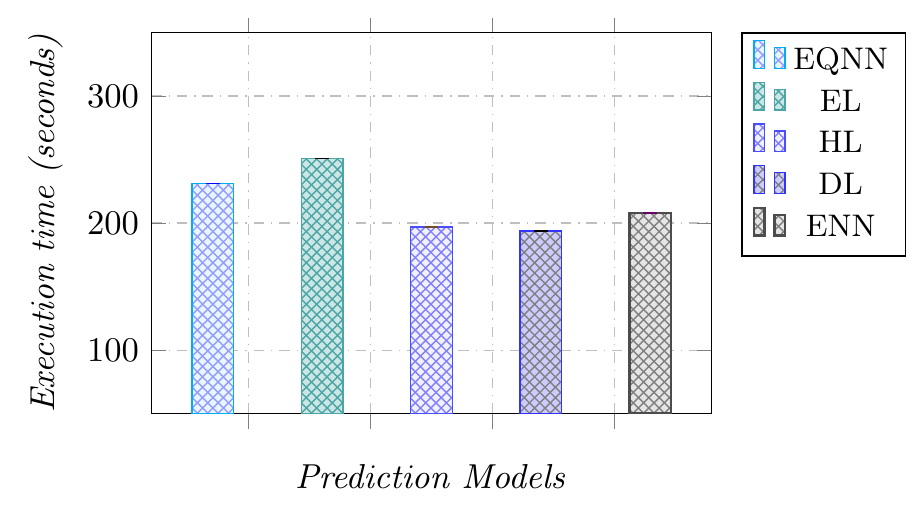}}\hfill
	\subfigure[GCD-Memory traces  ]{\includegraphics[width=.315\textwidth]{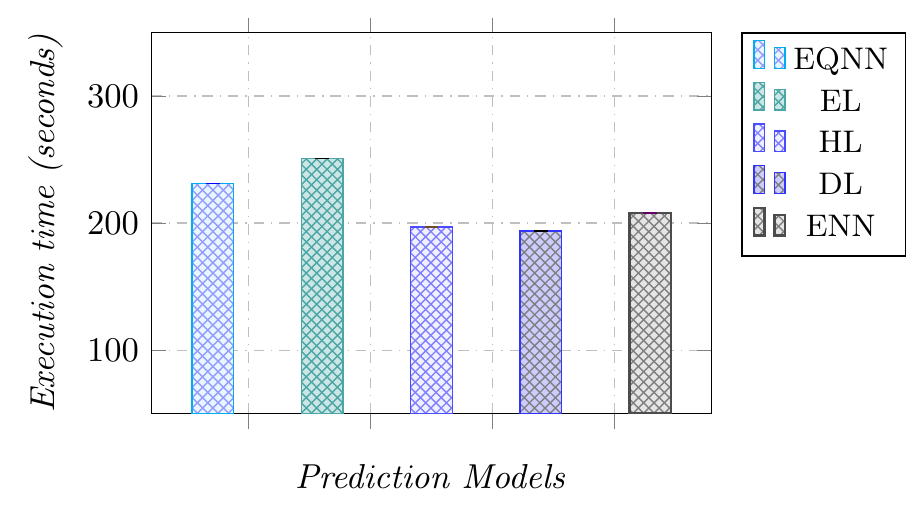}}\hfill
	\subfigure[PlanetLab CPU traces ]{\includegraphics[width=.315\textwidth]{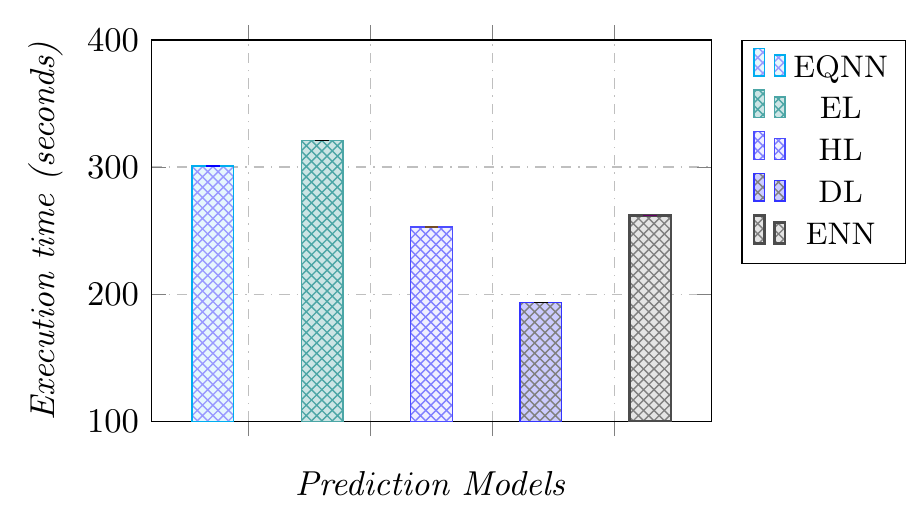}}
	
	\caption{Training time consumption }
	\label{fig:time}
	
\end{figure*}

The comparison of training-time elapsed during learning process of the various prediction models over distinct workload traces for the prediction window-size of 30 minutes is illustrated in Fig. \ref{fig:time}. HL and DL based models consumed similar time of training which is least among the training time of all the prediction models which is due to the usage of Gradient descent and Adam optimizer based training algorithms. While the time elapsed in the training of ENN, EL, and EQNN  is longer by reason of usage of evolutionary optimization during learning process. The training time for EL is higher than EQNN and ENN due to the engagement of multiple base prediction models which are simultaneously during the learning process. Contrary to this, the  single network based prediction models are used during the learning process of EQNN and ENN, where EQNN consume more time than ENN.     The reason behind this is the employment of Qubits and Quantum mechanics based network weight optimization process which uses highly complex computation dealing with complex numbers required for the  generation and updation of qubit-based network weights. This discussion concludes the trend for training time: DL $<$ HL$<$  ENN $<$ EQNN $<$ EL. However, the efficiency and applicability of the various prediction models is not affected because the training is a periodic task and can be executed in parallel on the servers equipped with enough resources. 

\subsubsection{Absolute Error Frequency } 

\begin{figure*}[!htbp]
	
	\centering
	
	\subfigure[GCD-CPU traces]{\includegraphics[width=.315\textwidth]{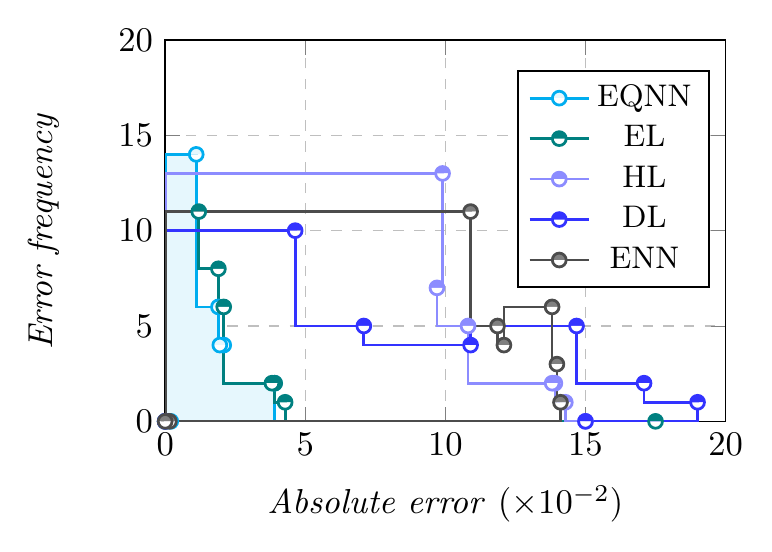}}\hfill
	\subfigure[GCD-Memory traces  ]{\includegraphics[width=.315\textwidth]{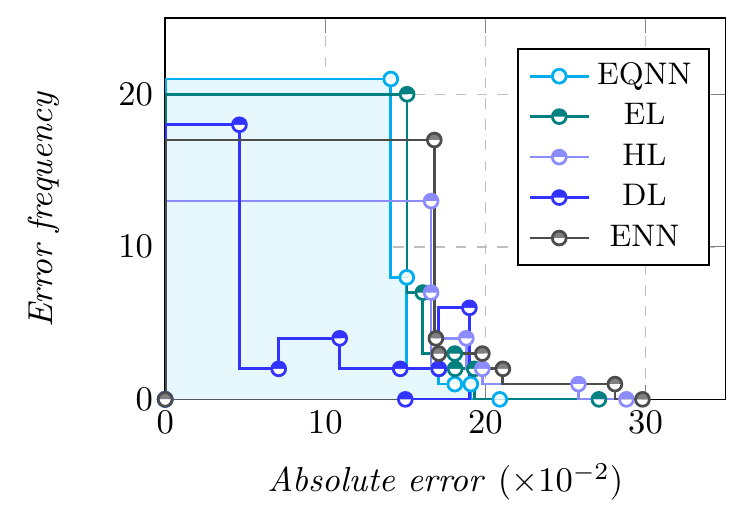}}\hfill
	\subfigure[PlanetLab CPU traces ]{\includegraphics[width=.315\textwidth]{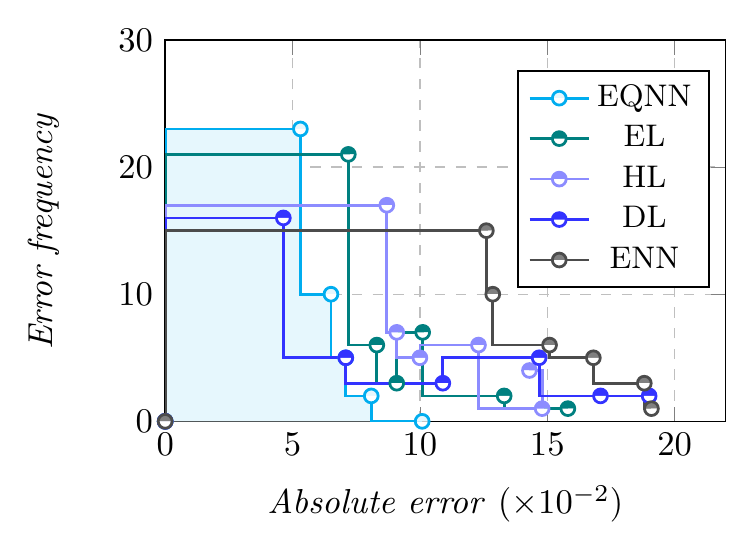}}
	
	\caption{Absolute Error Frequency }
	\label{fig:Aef}
	
\end{figure*}

The prediction error achieved for the various comparative models is measured and analysed by evaluating absolute prediction error and comparing its frequency for all three  workloads. Fig. \ref{fig:Aef} compares the frequency of absolute error (\textit{Actual value}-\textit{Predicted value}), where EQNN yields least error for the majority of the cloud workloads as compared with the other four types of prediction models.  The high frequency of absolute error for a prediction model indicates the consistent potency and stable tendency for yielding maximum prediction accuracy. The absolute error frequency observed for GCD-CPU traces (Fig. \ref{fig:Aef}a), GCD-Memory (Fig. \ref{fig:Aef}b), and  PL-CPU traces (Fig. \ref{fig:Aef}c) follows a common trend:   ENN $<$ DL $<$ HL $<$ EL $<$ EQNN. The reason behind such a trend is that EQNN employed Qubits population along with evolutionary optimization to allow an improved intuitive learning of patterns which concedes effective learning of extensive range of dynamic workload patterns with optimum accuracy. EL follows EQNN because of the involvement of the learning capability of multiple base predictor models which precisely learns the relevant information from the varying types of workloads.  The other prediction models also show slightly lesser but acceptable frequency of prediction error which varies according to the learning capabilities and optimization algorithms involved in their learning process.

\subsection{Trade-offs and Discussion}
All the machine learning algorithms have some trade-offs in relation to the adaptive prediction of extensive range of workloads. Likewise, the key difference among various ENNs-based prediction approaches is the evolutionary optimization algorithm applied for the learning process that directly impacts the performance of the prediction approach.  The different evolutionary optimization algorithms vary in exploration and exploitation methods involved in the population update process and control parameters tuning process. The evolutionary optimization approach having lesser number of hyperparameters for tuning is more preferable as compared to the one having higher number of tuning hyperparameters. For instance, a Blackhole learning algorithm is a parameterless algorithm  having lesser time and space complexity and predicts with higher accuracy than Differential Evolutionary algorithm that involves tuning of hyperparameters including crossover-rate, mutation-rate, learning rate etc. Though the Deep learning approach learns the natural variations of the data samples faster as compared to the evolutionary learning based feed-forward neural network models, they need larger number of data samples for training to estimate the output precisely. Also,deep learning approach having higher complexity computationally, requires expensive GPUs and high processing machines which scales up the cost of their applications. It has been observerd that deep learning algorithms perform better in integration with other machine learning approaches such as  random forest for feature extraction and compiles predicted output with the cooperation of the other classification approaches.  On the other hand, hybrid  and ensemle learning approaches involve combined operation of multiple machine learning algorithms consume higher space and time complexity over the single unit machine learning approaches. Undoubtedly, they adapt to the unseen data and extensive range of workloads with higher efficiency over deep and evolutionary learning approaches because of inclusion of several machine learning approaches at  a common platform. Among the hybrid and ensemble learning approaches, the ensemble approach is more adaptable as it considers prediction output from all the considered machine learning algorithms and applies weight optimization for selection of the predicted outcome associated to these learning algorithms while  generating the final output. The quantum neural network based learning aproach is most efficient among all the discussed approaches which is validated from the experiments for the accurate prediction of varying workloads. In QNN, the usage of qubits derived from complex numbers having higher diversity over real-numbered network weight values, enables to generate more intuitive pattern and learning of complex relations and helps to predict the output with higher accuracy. 

\par {Finally and most important, the diversity of cloud services such as IaaS, PaaS, SaaS, FaaS, etc., has a significant impact while deciding the heterogeneity of approaching workloads, and cloud service provider is bound to provide seamless quality and capacity of resources.  There is no perfect guideline to select the best model for a particular cloud service because the resource demands vary dynamically for all the cloud service models. Also, the cloud workloads vary because of various features including resource capacity (viz., CPU, memory, bandwidth, etc.) utilization; priority constraints such as deadline of execution; cost of execution; the amount of variability in the number of job requests over a time period, etc. However, these features (except the amount of variability in the number of job requests over a time period) do not have significant impact on the learning capability of different prediction models because the corresponding data samples are generated periodically, as event recordings for every type of workload, reporting all the relevant information over a timestamp. These data samples are used for training prediction models which show varying accuracy and training computation cost for the estimated workload over the same time period. However, the prediction models can be selected based on the priorities of constraints, such as; for promising high availability and deadline sensitivity, high accuracy prediction models should be selected irrespective of high computation and training time. On the other hand, if there is a constraint of minimum execution cost, the prediction model with lesser computation and training cost is a better option to minimize the processing cost of the respective workload.  Further, the workload prediction model can be chosen depending on the diversity of the workloads such as highly random, periodically variable, randomly variable, uniformly or non-uniformly diversifying, etc. Based on the above discussion of the characteristics, design, and capability of the considered machine learning models, it can be postulated that EQNN, Hybrid, and Ensemble models are more suitable for highly random and diversified workloads as they are more capable of learning and handling highly variable and heterogeneous traffic data patterns that involve large amounts of dynamic data, multiple variables with complicated relationships, and even multi-step time series  traffic data.  While the Evolutionary learning model is preferably suitable for periodically and uniformly variable workload observations recorded sequentially over equal time.} 

\section{Conclusions and Future Directions}
This paper presented a comprehensive survey and performance evaluation based comparison of the machine learning based workload prediction models  for  resource distribution and managemment in cloud environments. The operational design, utility, motivation, and challenges of the workload prediction approach are  discussed.  Based on the differences in the  conceptual and operational characteristics  of various prediction models, a  classification and taxonomy of machine learning driven prediction models is presented. The leading prediction approaches respective to each prediction model is thoroughly discussed. Further, all the discussed prediction models are implemented on a common platform for an extensive investigation and comparison of the performance of these models. Based on the intensive study and performance evaluation, a trade-off  among these prediction models and their applicability are discussed to conclude the holistic study of the cloud workload prediction  models.  In future, Explainable Artificial Intelligence (XAI) approach can be utilized to build more robust workload prediction models with retraceable mechanism that will help characterize accuracy,  transparency, fairness, and prediction outcomes in AI-powered resource management.  Further, the efficiency  of QNN prediction models can be improved by optimizing the qubit network with a lightweight optimization algorithm and reducing its computational complexity.   



%

\ifCLASSOPTIONcompsoc
  \section*{Acknowledgments}
\else
  \section*{Acknowledgment}
\fi

This research is supported by the National Institute of Technology, Kurukshetra, India and Goethe University, Frankfurt, and Austrian Science Fund (FWF) and the German Research Foundation (DFG), grant I 4800-N (ADVISE), 2020-2023. We thank the reviewers and editor for helping improve the manuscript.


\ifCLASSOPTIONcaptionsoff
  \newpage
\fi



%
\bibliographystyle{IEEEtran}
\bibliography{bibfile}

\begin{thebibliography}{10}
\providecommand{\url}[1]{#1}
\csname url@samestyle\endcsname
\providecommand{\newblock}{\relax}
\providecommand{\bibinfo}[2]{#2}
\providecommand{\BIBentrySTDinterwordspacing}{\spaceskip=0pt\relax}
\providecommand{\BIBentryALTinterwordstretchfactor}{4}
\providecommand{\BIBentryALTinterwordspacing}{\spaceskip=\fontdimen2\font plus
\BIBentryALTinterwordstretchfactor\fontdimen3\font minus
  \fontdimen4\font\relax}
\providecommand{\BIBforeignlanguage}[2]{{%
\expandafter\ifx\csname l@#1\endcsname\relax
\typeout{** WARNING: IEEEtran.bst: No hyphenation pattern has been}%
\typeout{** loaded for the language `#1'. Using the pattern for}%
\typeout{** the default language instead.}%
\else
\language=\csname l@#1\endcsname
\fi
#2}}
\providecommand{\BIBdecl}{\relax}
\BIBdecl

\bibitem{saxena2021op}
D.~Saxena, A.~K. Singh, and R.~Buyya, ``\uppercase{OP-MLB}: An online vm
  prediction based multi-objective load balancing framework for resource
  management at cloud datacenter,'' \emph{IEEE Trans. on Cloud Comp.}, 2021.

\bibitem{saxena2021secure}
D.~Saxena, I.~Gupta, J.~Kumar, A.~K. Singh, and X.~Wen, ``A secure and
  multiobjective virtual machine placement framework for cloud data center,''
  \emph{IEEE Systems Journal}, 2021.

\bibitem{mishra2022linking}
R.~Mishra, R.~Singh, and T.~Papadopoulos, ``Linking digital orientation and
  data-driven innovations: A sap-lap linkages framework and research
  propositions,'' \emph{IEEE Trans. on Engineering Management}, 2022.

\bibitem{trend2022cloud}
``Cloud comp. market size, share and trends analysis report by service iaas,
  paas, saas, by deployment public, private, hybrid, by enterprise size, by end
  use.''\hskip 1em plus 0.5em minus 0.4em\relax Segment Forecasts, 2022-2030.

\bibitem{ren2022machine}
Z.~Ren, J.~Wan, and P.~Deng, ``Machine-learning-driven digital twin for
  lifecycle management of complex equipment,'' \emph{IEEE Trans. on Emerg.
  Topics in Comp.}, 2022.

\bibitem{saxena2023sustainable}
D.~Saxena, A.~K. Singh, C.-N. Lee, and R.~Buyya, ``A sustainable and secure
  load management model for green cloud data centres,'' \emph{Scientific
  Reports}, 2023.

\bibitem{song2013adaptive}
W.~Song, Z.~Xiao, Q.~Chen, and H.~Luo, ``Adaptive resource provisioning for the
  cloud using online bin packing,'' \emph{IEEE Trans. on Computers}, vol.~63,
  no.~11, pp. 2647--2660, 2013.

\bibitem{saxena2020security}
D.~Saxena and A.~Singh, ``Security embedded dynamic resource allocation model
  for cloud data centre,'' \emph{Elec. Lttr.}, vol.~56, no.~20, pp. 1062--1065,
  2020.

\bibitem{saxena2021osc}
D.~Saxena and A.~K. Singh, ``Osc-mc: Online secure communication model for
  cloud environment,'' \emph{IEEE Comms. Lttr.}, vol.~25, no.~9, pp.
  2844--2848, 2021.

\bibitem{saxena2022ofp}
------, ``Ofp-tm: an online vm failure prediction and tolerance model towards
  high availability of cloud computing environments,'' \emph{The Journal of
  Supercomputing}, vol.~78, no.~6, pp. 8003--8024, 2022.

\bibitem{saxena2022high}
------, ``A high availability management model based on vm significance ranking
  and resource estimation for cloud applications,'' \emph{IEEE Trans. on
  Services Comp.}, 2022.

\bibitem{saxena2021workload}
------, ``Workload forecasting and resource management models based on machine
  learning for cloud computing environments,'' \emph{arXiv preprint
  arXiv:2106.15112}, 2021.

\bibitem{gupta2022quantum}
R.~Gupta, D.~Saxena, I.~Gupta, A.~Makkar, and A.~K. Singh, ``Quantum machine
  learning driven malicious user prediction for cloud network communications,''
  \emph{IEEE Netw. Lttr.}, 2022.

\bibitem{wang2022truthful}
X.~Wang, L.~Ma, X.~Wang, Y.~Shi, B.~Yi, and M.~Huang, ``Truthful vnfi
  procurement mechanisms with flexible resource provisioning in nfv markets,''
  \emph{IEEE Trans. on Cloud Comp.}, 2022.

\bibitem{saxena2020communication}
D.~Saxena and A.~K. Singh, ``Communication cost aware resource efficient load
  balancing (care-lb) framework for cloud datacenter,'' \emph{Recent Advances
  in Computer Science and Communications}, vol.~12, pp. 1--00, 2020.

\bibitem{singh2021cryptography}
A.~K. Singh and D.~Saxena, ``A cryptography and machine learning based
  authentication for secure data-sharing in federated cloud services
  environment,'' \emph{Journal of Applied Security Research}, pp. 1--24, 2021.

\bibitem{saxena2022intelligent}
D.~Saxena and A.~K. Singh, ``An intelligent traffic entropy learning-based load
  management model for cloud networks,'' \emph{IEEE Netw. Lttr.}, vol.~4,
  no.~2, pp. 59--63, 2022.

\bibitem{xie2022random}
Y.~Xie, L.~Pan, S.~Yang, and S.~Liu, ``A random online algorithm for reselling
  reserved iaas instances in amazon's cloud marketplace,'' \emph{IEEE Trans. on
  Network Science and Engineering}, 2022.

\bibitem{bi2019temporal}
J.~Bi, H.~Yuan, and M.~Zhou, ``Temporal prediction of multiapplication
  consolidated workloads in distributed clouds,'' \emph{IEEE Trans. on
  Automation Science and Engineering}, 2019.

\bibitem{kabir2021uncertainty}
H.~D. Kabir, A.~Khosravi, S.~K. Mondal, M.~Rahman, S.~Nahavandi, and R.~Buyya,
  ``Uncertainty-aware decisions in cloud computing: Foundations and future
  directions,'' \emph{ACM Comp. Surveys (CSUR)}, vol.~54, no.~4, pp. 1--30,
  2021.

\bibitem{saxena2020proactive}
D.~Saxena and A.~K. Singh, ``A proactive autoscaling and energy-efficient vm
  allocation framework using online multi-resource neural network for cloud
  data center,'' \emph{Neurocomputing}, 2020.

\bibitem{saxena2022fault}
D.~Saxena, I.~Gupta, A.~K. Singh, and C.-N. Lee, ``A fault tolerant elastic
  resource management framework towards high availability of cloud services,''
  \emph{IEEE Trans. on Network and Service Management}, 2022.

\bibitem{saxena2022intelligent1}
D.~Saxena and A.~K. Singh, ``an intelligent security centered
  resource-efficient resource management model for cloud computing
  environments,'' \emph{arXiv preprint arXiv:2210.16602}, 2022.

\bibitem{griner2021cerberus}
C.~Griner, J.~Zerwas, A.~Blenk, M.~Ghobadi, S.~Schmid, and C.~Avin, ``Cerberus:
  The power of choices in datacenter topology design-a throughput
  perspective,'' \emph{Proceedings of the ACM on Measurement and Analysis of
  Comp. Systems}, vol.~5, no.~3, pp. 1--33, 2021.

\bibitem{kumar2018workload}
J.~Kumar and A.~K. Singh, ``Workload prediction in cloud using artificial
  neural network and adaptive differential evolution,'' \emph{Future Generation
  Computer Systems}, vol.~81, pp. 41--52, 2018.

\bibitem{kumar2020biphase}
J.~Kumar, D.~Saxena, A.~K. Singh, and A.~Mohan, ``Biphase adaptive
  learning-based neural network model for cloud datacenter workload
  forecasting,'' \emph{Soft Comp.}, pp. 1--18, 2020.

\bibitem{kumar2021self}
J.~Kumar, A.~K. Singh, and R.~Buyya, ``Self directed learning based workload
  forecasting model for cloud resource management,'' \emph{Information
  Sciences}, vol. 543, pp. 345--366, 2021.

\bibitem{khorsand2018fahp}
R.~Khorsand, M.~Ghobaei-Arani, and M.~Ramezanpour, ``Fahp approach for
  autonomic resource provisioning of multitier applications in cloud computing
  environments,'' \emph{Software: Practice and Experience}, vol.~48, no.~12,
  pp. 2147--2173, 2018.

\bibitem{kumar2018long}
J.~Kumar, R.~Goomer, and A.~K. Singh, ``Long short term memory recurrent neural
  network (lstm-rnn) based workload forecasting model for cloud datacenters,''
  \emph{Procedia Computer Science}, vol. 125, pp. 676--682, 2018.

\bibitem{tang2019large}
X.~Tang, ``Large-scale computing systems workload prediction using parallel
  improved lstm neural network,'' \emph{IEEE Access}, vol.~7, pp.
  40\,525--40\,533, 2019.

\bibitem{gao2020task}
J.~Gao, H.~Wang, and H.~Shen, ``Task failure prediction in cloud data centers
  using deep learning,'' \emph{IEEE transactions on services computing}, 2020.

\bibitem{ruan2021workload}
L.~Ruan, Y.~Bai, S.~Li, S.~He, and L.~Xiao, ``Workload time series prediction
  in storage systems: a deep learning based approach,'' \emph{Cluster Comp.},
  pp. 1--11, 2021.

\bibitem{ruan2022cloud}
L.~Ruan, Y.~Bai, S.~Li, J.~Lv, T.~Zhang, L.~Xiao, H.~Fang, C.~Wang, and Y.~Xue,
  ``Cloud workload turning points prediction via cloud feature-enhanced deep
  learning,'' \emph{IEEE Trans. on Cloud Comp.}, 2022.

\bibitem{tuli2021start}
S.~Tuli, S.~S. Gill, P.~Garraghan, R.~Buyya, G.~Casale, and N.~Jennings,
  ``Start: Straggler prediction and mitigation for cloud comp. environments
  using encoder lstm networks,'' \emph{IEEE Trans. on Serv. Comp.}, 2021.

\bibitem{zhang2018efficient}
Q.~Zhang, L.~T. Yang, Z.~Yan, Z.~Chen, and P.~Li, ``An efficient deep learning
  model to predict cloud workload for industry informatics,'' \emph{IEEE Trans.
  on Industrial Informatics}, vol.~14, no.~7, pp. 3170--3178, 2018.

\bibitem{chen2019towards}
Z.~Chen, J.~Hu, G.~Min, A.~Y. Zomaya, and T.~El-Ghazawi, ``Towards accurate
  prediction for high-dimensional and highly-variable cloud workloads with deep
  learning,'' \emph{IEEE Trans. on Parallel and Distributed Systems}, vol.~31,
  no.~4, pp. 923--934, 2019.

\bibitem{peng2018multi}
C.~Peng, Y.~Li, Y.~Yu, Y.~Zhou, and S.~Du, ``Multi-step-ahead host load
  prediction with gru based encoder-decoder in cloud computing,'' in \emph{2018
  10th International Conference on Knowledge and Smart Technology (KST)}.\hskip
  1em plus 0.5em minus 0.4em\relax IEEE, 2018, pp. 186--191.

\bibitem{qiu2016deep}
F.~Qiu, B.~Zhang, and J.~Guo, ``A deep learning approach for vm workload
  prediction in the cloud,'' in \emph{2016 17th IEEE/ACIS Inter. Conf. on
  Software Engineering, Artificial Intelligence, Networking and
  Parallel/Distributed Comp. (SNPD)}.\hskip 1em plus 0.5em minus 0.4em\relax
  IEEE, 2016, pp. 319--324.

\bibitem{zhang2017resource}
W.~Zhang, P.~Duan, L.~T. Yang, F.~Xia, Z.~Li, Q.~Lu, W.~Gong, and S.~Yang,
  ``Resource requests prediction in the cloud computing environment with a deep
  belief network,'' \emph{Software: Practice and Experience}, vol.~47, no.~3,
  pp. 473--488, 2017.

\bibitem{wen2020cpu}
Y.~Wen, Y.~Wang, J.~Liu, B.~Cao, and Q.~Fu, ``Cpu usage prediction for cloud
  resource provisioning based on deep belief network and particle swarm
  optimization,'' \emph{Concurrency and Computation: Practice and Experience},
  vol.~32, no.~14, p. e5730, 2020.

\bibitem{xu2022esdnn}
M.~Xu, C.~Song, H.~Wu, S.~S. Gill, K.~Ye, and C.~Xu, ``Esdnn: Deep neural
  network based multivariate workload prediction approach in cloud
  environment,'' \emph{arXiv preprint arXiv:2203.02684}, 2022.

\bibitem{bhagtya2021workload}
P.~Bhagtya, S.~Raghavan, and K.~Chandraseakran, ``Workload classification in
  multi-vm cloud environment using deep neural network model,'' in
  \emph{Proceedings of the 36th Annual ACM Symposium on Applied Comp.}, 2021,
  pp. 79--82.

\bibitem{li2016learning}
Y.~Li, H.~Hu, Y.~Wen, and J.~Zhang, ``Learning-based power prediction for data
  centre operations via deep neural networks,'' in \emph{Proceedings of the 5th
  International Workshop on Energy Efficient Data Centres}, 2016, pp. 1--10.

\bibitem{bi2019deep}
J.~Bi, S.~Li, H.~Yuan, Z.~Zhao, and H.~Liu, ``Deep neural networks for
  predicting task time series in cloud computing systems,'' in \emph{2019 IEEE
  16th International Conference on Networking, Sensing and Control
  (ICNSC)}.\hskip 1em plus 0.5em minus 0.4em\relax IEEE, 2019, pp. 86--91.

\bibitem{kardani2020adrl}
S.~Kardani-Moghaddam, R.~Buyya, and K.~Ramamohanarao, ``Adrl: A hybrid
  anomaly-aware deep reinforcement learning-based resource scaling in clouds,''
  \emph{IEEE Trans. on Parallel and Distributed Systems}, vol.~32, no.~3, pp.
  514--526, 2020.

\bibitem{karim2021bhyprec}
M.~E. Karim, M.~M.~S. Maswood, S.~Das, and A.~G. Alharbi, ``Bhyprec: A novel
  bi-lstm based hybrid recurrent neural network model to predict the cpu
  workload of cloud virtual machine,'' \emph{IEEE Access}, vol.~9, pp.
  131\,476--131\,495, 2021.

\bibitem{bi2021integrated}
J.~Bi, S.~Li, H.~Yuan, and M.~Zhou, ``Integrated deep learning method for
  workload and resource prediction in cloud systems,'' \emph{Neurocomputing},
  vol. 424, pp. 35--48, 2021.

\bibitem{chen2015self}
Z.~Chen, Y.~Zhu, Y.~Di, and S.~Feng, ``Self-adaptive prediction of cloud
  resource demands using ensemble model and subtractive-fuzzy clustering based
  fuzzy neural network,'' \emph{Computational intelligence and neuroscience},
  vol. 2015, 2015.

\bibitem{liu2017adaptive}
C.~Liu, C.~Liu, Y.~Shang, S.~Chen, B.~Cheng, and J.~Chen, ``An adaptive
  prediction approach based on workload pattern discrimination in the cloud,''
  \emph{Journal of Network and Computer Applications}, vol.~80, pp. 35--44,
  2017.

\bibitem{shuvo2020lsru}
M.~N.~H. Shuvo, M.~M.~S. Maswood, and A.~G. Alharbi, ``Lsru: A novel deep
  learning based hybrid method to predict the workload of virtual machines in
  cloud data center,'' in \emph{2020 IEEE Region 10 Symposium (TENSYMP)}.\hskip
  1em plus 0.5em minus 0.4em\relax IEEE, 2020, pp. 1604--1607.

\bibitem{singh2014ensemble}
N.~Singh and S.~Rao, ``Ensemble learning for large-scale workload prediction,''
  \emph{IEEE Trans. on Emg. Topics in Comp.}, vol.~2, no.~2, pp. 149--165,
  2014.

\bibitem{feng2022fast}
B.~Feng, Z.~Ding, and C.~Jiang, ``Fast: A forecasting model with adaptive
  sliding window and time locality integration for dynamic cloud workloads,''
  \emph{IEEE Trans. on Serv. Comp.}, 2022.

\bibitem{kim2020forecasting}
I.~K. Kim, W.~Wang, Y.~Qi, and M.~Humphrey, ``Forecasting cloud application
  workloads with cloudinsight for predictive resource management,'' \emph{IEEE
  Trans. on Cloud Comp.}, 2020.

\bibitem{iqbal2019adaptive}
W.~Iqbal, J.~L. Berral, A.~Erradi, D.~Carrera \emph{et~al.}, ``Adaptive
  prediction models for data center resources utilization estimation,''
  \emph{IEEE Trans. on Network and Service Management}, vol.~16, no.~4, pp.
  1681--1693, 2019.

\bibitem{kumar2020ensemble}
J.~Kumar, A.~K. Singh, and R.~Buyya, ``Ensemble learning based predictive
  framework for virtual machine resource request prediction,''
  \emph{Neurocomputing}, vol. 397, pp. 20--30, 2020.

\bibitem{singh2021quantum}
A.~K. Singh, D.~Saxena, J.~Kumar, and V.~Gupta, ``A quantum approach towards
  the adaptive prediction of cloud workloads,'' \emph{IEEE Trans. on Parallel
  and Distributed Systems}, 2021.

\bibitem{prevost2011prediction}
J.~J. Prevost, K.~Nagothu, B.~Kelley, and M.~Jamshidi, ``Prediction of cloud
  data center networks loads using stochastic and neural models,'' in
  \emph{2011 6th International Conference on System of Systems
  Engineering}.\hskip 1em plus 0.5em minus 0.4em\relax IEEE, 2011, pp.
  276--281.

\bibitem{saxena2020auto}
D.~Saxena and A.~K. Singh, ``Auto-adaptive learning-based workload forecasting
  in dynamic cloud environment,'' \emph{International Journal of Computers and
  Applications}, pp. 1--11, 2020.

\bibitem{kumar2016dynamic}
J.~Kumar and A.~K. Singh, ``Dynamic resource scaling in cloud using neural
  network and black hole algorithm,'' in \emph{2016 Fifth International
  Conference on Eco-friendly Comp. and Communication Systems (ICECCS)}.\hskip
  1em plus 0.5em minus 0.4em\relax IEEE, 2016, pp. 63--67.

\bibitem{zhang2015towards}
W.~Zhang and P.~Duan, ``Towards a deep belief network-based cloud resource
  demanding prediction,'' in \emph{2015 IEEE 12th Intl Conf on Ubiquitous
  Intelligence and Comp. and 2015 IEEE 12th Intl Conf on Autonomic and Trusted
  Comp. and 2015 IEEE 15th Intl Conf on Scalable Comp. and Communications and
  Its Associated Workshops (UIC-ATC-ScalCom)}.\hskip 1em plus 0.5em minus
  0.4em\relax IEEE, 2015, pp. 1043--1048.

\bibitem{chen2019prediction}
D.~Chen, X.~Zhang, L.~L. Wang, and Z.~Han, ``Prediction of cloud resources
  demand based on hierarchical pythagorean fuzzy deep neural network,''
  \emph{IEEE Trans. on Serv. Comp.}, 2019.

\bibitem{qazi2018cloud}
K.~Qazi and I.~Aizenberg, ``Cloud datacenter workload prediction using
  complex-valued neural networks,'' in \emph{2018 IEEE Second International
  Conference on Data Stream Mining \& Processing (DSMP)}.\hskip 1em plus 0.5em
  minus 0.4em\relax IEEE, 2018, pp. 315--321.

\bibitem{Reiss2011}
J.~L.~H. C.~Reiss, J.~Wilkes, ``Google–cluster traces:format+schema,''
  \emph{Google Inc., White Paper}, 2011.

\bibitem{beloglazov2012optimal}
A.~Beloglazov and R.~Buyya, ``Optimal online deterministic algorithms and
  adaptive heuristics for energy and performance efficient dynamic
  consolidation of virtual machines in cloud data centers,'' \emph{Concurrency
  and Computation: Practice and Experience}, vol.~24, no.~13, pp. 1397--1420,
  2012.

\end{thebibliography}

%
\vskip 0pt plus -1fil
\begin{IEEEbiography} [{\includegraphics[width=0.8\linewidth]{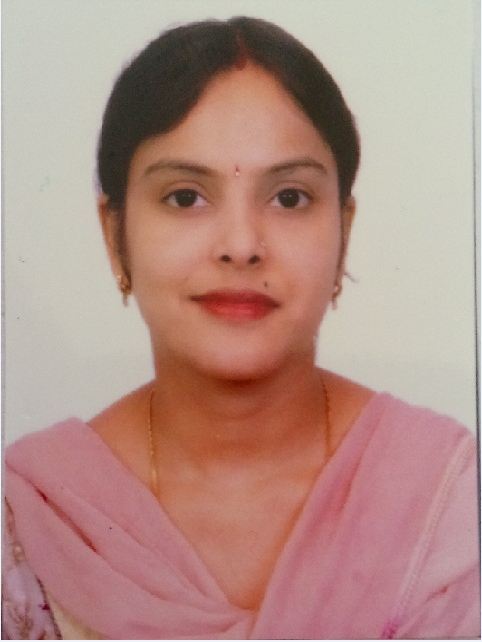}}]{Deepika Saxena} is a Postdoctoral Research
	Associate at Goethe University, Frankfurt. She
	earned her Ph. D. degree from the Department
	of Computer Applications, National Institute of
	Technology (NIT), Kurukshetra, India. She received her M.Tech (CSE) degree from Kurukshetra University Kurukshetra, India in 2014.
	Her major research interests are Neural Networks, Evolutionary Algorithms, Resource Management, and Security in Cloud Computing.
	
\end{IEEEbiography}
\vskip 0pt plus -1fil 
\begin{IEEEbiography}[{\includegraphics[width=0.8\linewidth]{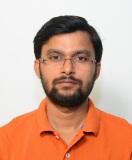}}]{Jitendra Kumar} is an Assistant Professor in the Department of Computer Applications, National Institute of Technology Tiruchirappalli, India. He earned his doctorate from the National Institute of Technology Kurukshetra, India in 2019. His current research interests include Cloud Computing, Machine Learning, Data Analytics, Parallel Processing.
\end{IEEEbiography}
\vskip 0pt plus -1fil 
\begin{IEEEbiography}[{\includegraphics[width=0.8\linewidth]{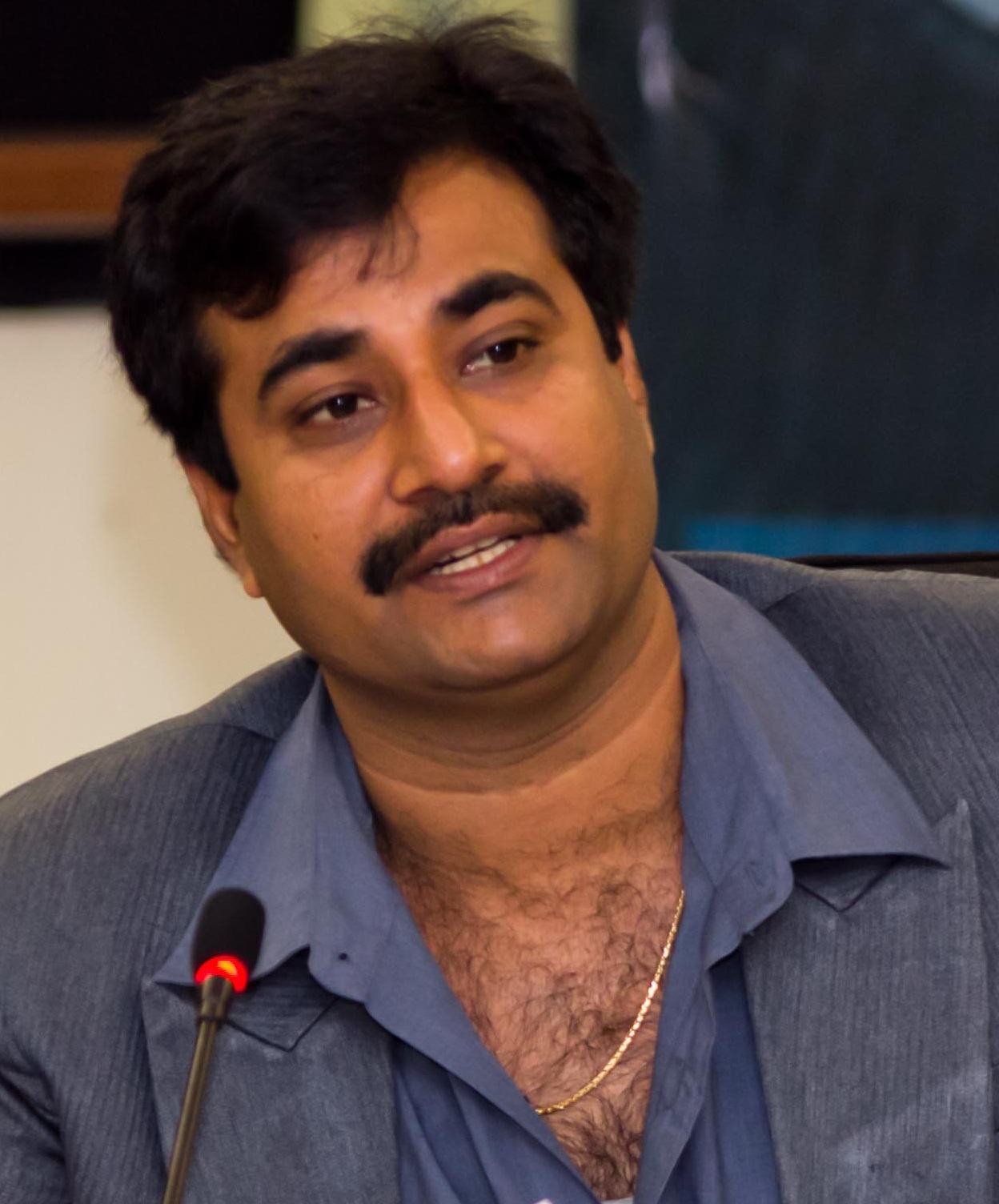}}]{Ashutosh Kumar Singh}
	is working as a Professor in the Department of Computer Applications, National Institute of Technology Kurukshetra, India. He has more than 20 years research in various Universities of the India, UK, and Malaysia. He received his PhD from Indian Institute of Technology, BHU, India and Post Doc from Department of Computer Science, University of Bristol, UK. He is also Charted Engineer from UK. His research area includes Verification, Synthesis, Design and Testing of Digital Circuits, Data Science, Cloud Computing, Machine Learning, Security, Big Data. 
\end{IEEEbiography}
\vskip 0pt plus -1fil 
\begin{IEEEbiography}[{\includegraphics[width=0.8\linewidth]{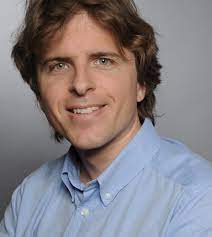}}]{Stefan Schmid} is a Professor at the Technical University of Berlin, Germany, working part-time for the Fraunhofer Institute for Secure Information Technology (SIT) in Germany as well as for the Faculty of Computer Science, the University of Vienna in Austria. MSc and Ph.D. at ETH Zurich, Postdoc at TU Munich and the University of Paderborn, Senior Research Scientist at T-Labs in Berlin, Associate Professor at Aalborg University, Denmark, and Full Professor at the University of Vienna, Austria. He received the IEEE Communications Society ITC Early Career Award 2016 and an ERC Consolidator Grant 2019.
\end{IEEEbiography}



\end{document}